\documentclass[
amsmath,
prd,nofootinbib,floatfix,11pt,
]{revtex4}

\usepackage{placeins}
\usepackage{afterpage}

\usepackage{axodraw2}
\usepackage{graphicx}
\usepackage{setspace}
\usepackage{bm}

\newcommand\beq{\begin{eqnarray}}
\newcommand\eeq{\end{eqnarray}}
\def\lsim{\mathrel{\rlap{\lower4pt\hbox{$\sim$}}
    \raise1pt\hbox{$<$}}}                
\def\gsim{\mathrel{\rlap{\lower4pt\hbox{$\sim$}}
    \raise1pt\hbox{$>$}}}            

\allowdisplaybreaks
\newcommand\lnbar{\overline{\ln}}
\newcommand\MSbar{$\overline{\rm{MS}}$ }
\newcommand\Aeps{A_{\epsilon}}
\newcommand\Aepstwo{A_{\epsilon^2}}
\newcommand\Lsthree{{\rm Ls_3}}
\newcommand\Lspfour{{\rm Ls'_4}}

\def\cI{c_{I}}
\def\conFbar{c_{F}}
\def\conG{c_{G}}
\def\conHfour{c'_{H}}
\def\conHfive{c''_{H}}
\def\conHsix{c_{H}}

\begin{document}

\renewcommand{\theequation}{\arabic{section}.\arabic{equation}}
\renewcommand{\thefigure}{\arabic{section}.\arabic{figure}}
\renewcommand{\thetable}{\arabic{section}.\arabic{table}}

\title{\Large \baselineskip=20pt 
Renormalized $\epsilon$-finite master integrals and their virtues:\\
the three-loop self energy case}
\author{Stephen P.~Martin}
\affiliation{\mbox{\it Department of Physics, Northern Illinois University, DeKalb IL 60115}}

\begin{abstract}\normalsize \baselineskip=15pt 
Loop diagram calculations typically rely on reduction to a finite set of master integrals in $4 - 2 \epsilon$ dimensions. It has been shown that for any problem, the masters can be chosen so that their coefficients are finite as $\epsilon \rightarrow 0$. I propose a definition of renormalized $\epsilon$-finite master integrals, which incorporate ultraviolet divergence subtractions in a specific way. A key advantage of this choice is that in expressions for physical observables, expansions to positive powers in $\epsilon$ are never needed. As an example, I provide the subtractions for general three-loop self-energy integrals. The differential equations method is used to compute numerically the renormalized $\epsilon$-finite master integrals for arbitrary external momentum invariant, in special cases with internal masses equal to a single scale or zero. These include the ones necessary for the three-loop QCD corrections to the self-energies of the $W$, $Z$, and Higgs bosons. In principle, the same method should provide for numerical computation of general three-loop self energies with any masses.
\end{abstract}

\maketitle

\vspace{-0.2in}

\baselineskip=14.5pt

\tableofcontents

\baselineskip=14.7pt
\newpage

\section{Introduction\label{sec:intro}}
\setcounter{equation}{0}
\setcounter{figure}{0}
\setcounter{table}{0} 
\setcounter{footnote}{1}

Precision calculations of radiative corrections in relativistic quantum field theory in the modern era almost always make use of dimensional regularization \cite{Bollini:1972ui,Bollini:1972bi,Ashmore:1972uj,Cicuta:1972jf,tHooft:1972tcz,tHooft:1973mfk}
to
\beq
d = 4 - 2 \epsilon
\eeq
dimensions
in order to deal with ultraviolet (UV)  and sometimes infrared (IR) divergences. The integration by parts (IBP) method  \cite{Tkachov:1981wb,Chetyrkin:1981qh} can then be used to reduce the expressions to linear combinations of so-called master integrals, with coefficients that are rational expressions in $\epsilon$, the propagator squared masses, and the external momentum invariants. There are an infinite number of IBP relations, but only a finite number \cite{Smirnov:2010hn} of master integrals are needed to express the results for any given problem. In principle, the method of ref.~\cite{Laporta:2000dsw} can always be used to solve the IBP relations. However, in practice the reduction process can have formidable memory and computing time requirements, which has prompted the development of various advanced algorithms and computer codes 
\cite{Anastasiou:2004vj,Smirnov:2008iw,Smirnov:2014hma,Smirnov:2019qkx,Studerus:2009ye,vonManteuffel:2012np,Lee:2012cn,Lee:2013mka,vonManteuffel:2014ixa,Georgoudis:2016wff,Maierhofer:2017gsa,Maierhofer:2018gpa,Klappert:2020nbg}
to solve the problem.
  
The choice of master integrals is not unique, and there are at least three distinct criteria one might use to choose them. One goal might be to simplify as much as possible the task of reduction of a general integral to the masters. A second possible criteria could be to simplify the analytic or numerical calculation of the master integrals themselves. A third criteria might be to simplify as much as possible the presentation of results for physical observables. These criteria need not coincide, and can naturally lead to quite different choices for the master integrals. Some proposals for how to choose the master integrals in various contexts are given in 
refs.~\cite{Chetyrkin:2006dh,Faisst:2006sr,Gluza:2010ws,Schabinger:2011dz,Henn:2013pwa,Henn:2014qga,Lee:2014ioa,Meyer:2016slj,vonManteuffel:2017myy,Lee:2019wwn,Smirnov:2020quc,Usovitsch:2020jrk}.

In the present paper, I will be interested in the specific goal of making the presentation of physical observables in terms of the master integrals as simple as possible. First, one often has to deal with the issue of ``spurious" poles in $\epsilon$, which occur not in the master integrals themselves but in the coefficients multiplying the master integrals in some physical quantity of interest. These can be a quite common occurrence when propagator masses vanish, and leads to the following problem. If the coefficient of a master integral has a pole $1/\epsilon^n$, then one will need the expansion of the master integral itself up to order $\epsilon^n$ in order to obtain a correct expression in the $\epsilon \rightarrow 0$ limit. 

Fortunately, it was shown by Chetyrkin, Faisst, Sturm, and Tentyukov in ref.~\cite{Chetyrkin:2006dh} that for any problem one can always make a choice of the master integrals, called an $\epsilon$-finite basis, so that the coefficients multiplying them are finite as $\epsilon \rightarrow 0$. The choice of an $\epsilon$-finite basis\footnote{Note that it is only the coefficients that are finite as $\epsilon \rightarrow 0$, not the master integrals. If the propagator squared masses are all non-zero and generic, then any basis without explicit factors of $\epsilon$ is $\epsilon$-finite. Also, in practice, the ``basis" chosen might actually be over-complete, either because not all linear relations between them are known, or because imposing some of the known linear relations would cause unwelcome complexity in coefficients.} is neither unique nor obvious in general, but the existence proof also provides a simple algorithm for its construction. Moreover, for a given diagram topology class, this property of $\epsilon$-finiteness is independent of the physical observable being calculated.

However, even after choosing an $\epsilon$-finite basis for a given fixed loop order,  there is another problem to be considered. Suppose one is doing a calculation at $l$-loop order in perturbation theory, using master integrals at every loop order $k$, with $1 \leq k \leq l$. Then, when computing a renormalized quantity, each $k$-loop order master could be multiplied by an $(l-k)$-loop-order counterterm, and can also occur in factorized integrals multiplied by other master integrals whose loop order totals $l-k$.  In both cases, $k$-loop order master integrals will be multiplied by  poles as severe as $1/\epsilon^{l-k}$. This would seem to suggest that for an $l$-loop order calculation,  even with an $\epsilon$-finite basis, the expansion of masters of lower loop order $k$ will be needed for all positive powers up to $\epsilon^{l-k}$.
 
In this paper, I emphasize that the last problem is also avoided if one expresses results in terms of what I will call renormalized $\epsilon$-finite master integrals.
As explained in more detail in the next section, these are obtained from the 
$\epsilon$-finite masters by subtracting UV sub-divergences in a specific way, and then taking the limit as $\epsilon \rightarrow 0$. 
The key point is that when presenting results for the calculations of renormalized observables, by organizing the results in terms of
renormalized $\epsilon$-finite masters, it is never necessary to expand to positive
powers in $\epsilon$. This remains true even if the calculation is later extended to an arbitrary higher loop order.
A heuristic justification for why this pleasant feature is not completely unexpected is that in the calculation of renormalized physical observables, one could in principle employ some other regulator not based on dimensional continuation at all, in which case there would be no essential reason for the appearance of higher moments of the integrals continued away from $d=4$.

In this sense, the renormalized $\epsilon$-finite masters provide an optimal
way
of expressing and numerically computing physical results, since the components with positive
powers of $\epsilon$ do not appear and will never be needed. The essential reason for this is that the necessary renormalization of UV divergences automatically works together with the counterterms included within the definitions of the masters themselves, while IR divergences and other kinetic singularities must cancel if the calculated quantity is indeed an observable. This  has already been verified for a variety of effective potential, tadpole, and self-energy calculations up to (now) three-loop order, as detailed below.

The rest of this paper is organized as follows. In the next section, I give a definition of renormalized $\epsilon$-finite master integrals. In section \ref{sec:SE}, I explicitly provide the necessary definitions for three-loop self-energy (and vacuum) functions, which are the focus of the rest of the paper. In section \ref{sec:zeromass}, I review the results for the case of internal propagators that are all massless, and in section \ref{sec:allmassive} for the case that all internal propagators have the same non-zero mass. Sections \ref{sec:odd} and \ref{sec:even} treat the case of integrals that arise in the three-loop QCD contributions to the self-energies of the $W$ boson and the $Z,H$ boson in the Standard Model, respectively. These have one non-zero propagator mass (that of the top quark) and other internal masses (for gluons and other quarks) vanishing. Those results, obtained in the pure \MSbar tadpole-free scheme, will appear in a separate paper \cite{SPMtoappear}. In each case, a method for straightforward numerical computation of the renormalized $\epsilon$-finite master integrals (valid even for the range of external momentum invariant such that the expansions around zero and infinite external momenta do not converge) is given, based on the differential equations method \cite{Kotikov:1990kg,Kotikov:1991hm,Ford:1991hw,Ford:1992pn,Bern:1993kr,Remiddi:1997ny,Caffo:1998du,Gehrmann:1999as,Caffo:2002ch,Caffo:2002wm,Caffo:2003ma,Martin:2003qz,Martin:2005qm,Martin:2016bgz}. Section \ref{sec:outlook} contains some concluding remarks.

\section{Renormalized $\epsilon$-finite master integrals \label{sec:renormintegrals}}
\setcounter{equation}{0}
\setcounter{figure}{0}
\setcounter{table}{0} 
\setcounter{footnote}{1}

Consider an $l$-loop scalar integral ${\bf I}$ in $d = 4 - 2\epsilon$ dimensions, 
which depends on some propagator squared masses and external momentum invariants. 
Suppose that ${\bf I}$ is a member of an $\epsilon$-finite basis, as in ref.~\cite{Chetyrkin:2006dh}.
Let us define the corresponding renormalized integral $I$ according to
\beq
I &=& \lim_{\epsilon \rightarrow 0} \left [
{\bf I} - \sum_{k=0}^l {\bf I}^{k,\,{\rm div}} \right ]
,
\label{eq:defIrenormalized}
\eeq
where the UV $k$-loop sub-divergences ${\bf I}^{k,\,{\rm div}}$ have been subtracted.
More specifically,
\beq
{\bf I}^{k,\,{\rm div}} &=&  \sum_{{\bf J}_k} {\bf J}_k
\sum_{n=1}^k \frac{1}{\epsilon^n} c^{(n)}_{{\bf J}_{k}}
,
\label{eq:defIcounterterms}
\eeq
where the ${\bf J}_{k}$ are the integrals obtained from ${\bf I}$ by collapsing UV-divergent $k$-loop sub-diagrams to a point and eliminating the corresponding momentum integrations. Thus, each ${\bf J}_{k}$ is an $(l-k)$-loop integral, and in particular
${\bf J}_l = 1$. The sum over ${\bf J}_k$ is obtained by considering all of the complementary collapsed $k$-loop sub-diagrams that contain UV poles. The counterterm coefficients $c^{(n)}_{{\bf J}_{k}}$ are polynomials in the propagator squared masses and the external momentum invariants, chosen so that $I$ is free of UV divergences. Here, the UV divergences are defined to be those obtained for generic propagator squared masses and external momentum invariants. All remaining poles in $\epsilon$ 
are called infrared (IR) here, although it might be more precise to say ``non-UV".\footnote{Integrals evaluated at thresholds can have non-UV poles in $\epsilon$ that are also not IR divergences but are treated in the same way.} In the self-energy and vacuum integral cases studied explicitly below, each renormalized $\epsilon$-finite master integral $I$ is well-defined and finite for $\epsilon \rightarrow 0$, and so is independent of $\epsilon$, but for more external legs it might be useful to keep remaining poles as $1/\epsilon_{\rm IR}^n$. 

One can also expand the original integral ${\bf I}$ in powers $\epsilon^n$, starting from the leading pole at $n=-l$:
\beq
{\bf I} = \sum_{n = -l}^\infty \epsilon^n I^{(n)}.
\eeq
However, I propose that 
physical (renormalized) results should always be presented in terms of the integrals $I$, and
not in terms of the integrals $I^{(0)}$, which are different except in the case that
${\bf I}$ is already finite. If one uses the $I^{(0)}$ integrals, then master integrals
found at lower loop order will have to be expanded to positive powers in $\epsilon$. Instead, organizing the results in terms of the integrals $I$ avoids this, and is most convenient for extensions of the calculation to higher orders.

Renormalized $\epsilon$-finite masters have already been defined exactly as above and employed in various self-energy calculations through two-loop order in refs.~\cite{Martin:2003it,Martin:2004kr,Martin:2005eg,Martin:2014cxa,Martin:2005ch,Martin:2016xsp,Martin:2015lxa,Martin:2015rea}, and in the calculation of the effective potential through three-loop order in refs.~\cite{Martin:2001vx,Martin:2018emo,Martin:2017lqn}. In those previous examples, the ``renormalized" part of the definition of the masters was paramount, ensuring that positive powers in $\epsilon$ for one-loop and two-loop masters were not needed. The $\epsilon$-finiteness did not really play a role, simply because IR divergences were regulated by giving small regulator masses to gauge bosons, Goldstone bosons, and chiral fermions, rather than giving them exactly zero mass from the start. In this paper, I will treat the case of self-energy functions and vacuum integrals up to three-loop order, with applications to QCD corrections to weak boson self-energies in which gluons and the quarks other than the top quark will be treated as exactly massless from the start. These results appear in a companion paper ref.~\cite{SPMtoappear}, and illustrate the thematic property that expansions of the one-loop, two-loop, and three-loop masters to positive powers in $\epsilon$ are never needed.

\baselineskip=15.pt

\section{Self-energy integrals\label{sec:SE}}
\setcounter{equation}{0}
\setcounter{figure}{0}
\setcounter{table}{0} 
\setcounter{footnote}{1}

\subsection{General conventions\label{subsec:generalcon}}

In this section I establish the notations and conventions to be used below. Momentum integrals are defined in terms of their Wick-rotated Euclidean versions in $d=4 - 2\epsilon$ dimensions. In diagrams below, each line carrying 4-momentum $k^\mu$ and with squared mass $x$ represents a propagator factor of $1/(k^2 + x)$, and the loop-momentum integration measure is
\beq
\int_k &\equiv& (16 \pi^2) \frac{\mu^{2 \epsilon}}{(2 \pi)^d} \int d^d k.
\eeq
The regularization scale $\mu$ is then traded for a scale $Q$ (equal to the renormalization scale if the \MSbar scheme \cite{Bardeen:1978yd,Braaten:1981dv} is adopted), according to
\beq
Q^2 &=& 4 \pi e^{-\gamma} \mu^2,
\eeq
in terms of the Euler constant $\gamma = 0.5772156649\ldots$. Now define
\beq
L_x \equiv \lnbar(x) \equiv \ln(x/Q^2) ,
\eeq
where the second notation was used in refs.~\cite{Martin:2003qz,Martin:2005qm,Martin:2016bgz} and the first notation will be used below. The external momentum invariant for self-energy functions is defined to be
\beq
s \equiv -p^2 + i \varepsilon
,
\eeq
with a Euclidean (or signature $-$$+$$+$$+$) metric,
so that
\beq
L_{-s} \equiv \lnbar(-s) = \lnbar(s) - i \pi
,
\eeq 
where the last equation holds for positive (physical) $s$.
Below, $s$ and $Q$ will always be suppressed as function arguments, 
because they are always the same for
all self-energy functions in a given expression or equation. 

\subsection{One-loop and two-loop self-energy integrals\label{subsec:oneSE}}

\begin{figure}[!t]
\begin{center}
\includegraphics[width=13cm,angle=0]{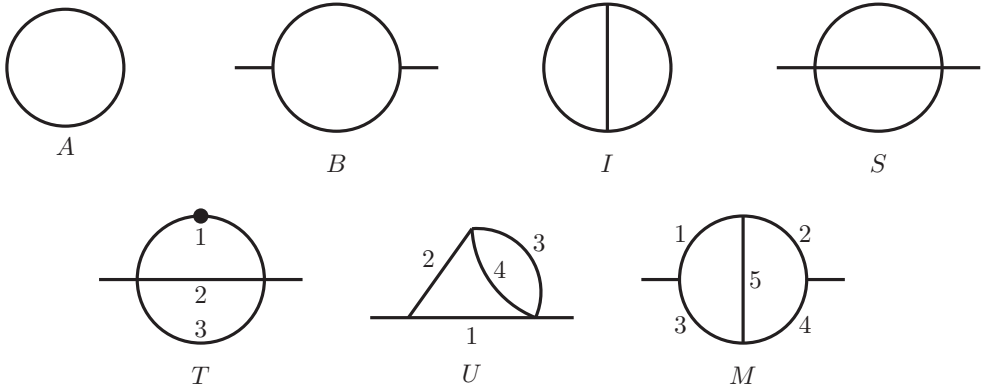}
\end{center}
\vspace{-0.4cm}
\begin{minipage}[]{0.95\linewidth}
\caption{\label{fig:onetwolooptopologies}
Topologies for one-loop and two-loop self-energy and vacuum master integrals in eqs.~(\ref{eq:defboldA})-(\ref{eq:defboldM}), following the same conventions and notations used in refs.~\cite{Martin:2003qz,Martin:2005qm,Martin:2016bgz}. The integer labels on the internal lines denote the ordering of internal propagator squared mass arguments.}
\end{minipage}
\end{figure}

The master integrals for one-loop and two-loop scalar self-energy integrals are as shown in Figure \ref{fig:onetwolooptopologies}, following the same notations and conventions as in refs.~\cite{Martin:2003qz,Martin:2005qm,Martin:2016bgz}.
Thus the (non-renormalized) master integrals at one loop are:
\beq
{\bf A}(x) &=& \int_k \frac{1}{k^2 + x},
\label{eq:defboldA}
\\
{\bf B}(x,y) &=& \int_k \frac{1}{[k^2 + x][(k-p)^2 + y]},
\eeq
and at two loops,
\beq
{\bf I}(x,y,z) &=& \int_k\int_q \frac{1}{[k^2 + x][q^2 + y][(k+q)^2 + z]},
\\
{\bf S}(x,y,z) &=& \int_k\int_q \frac{1}{[k^2 + x][q^2 + y][(k+q-p)^2 + z]},
\\
{\bf T}(x,y,z) &=& \int_k\int_q \frac{1}{[k^2 + x]^2\,[q^2 + y][(k+q-p)^2 + z]},
\\
{\bf U}(w,x,y,z) &=& \int_k\int_q \frac{1}{[k^2 + w][(k-p)^2 + x][q^2 + y][(k+q-p)^2 + z]},
\\
{\bf M}(v,w,x,y,z) &=& \int_k\int_q 
\frac{1}{[k^2 + v][q^2+w][(k-p)^2 + x][(q-p)^2 + y][(k-q)^2 + z]}.
\label{eq:defboldM}
\eeq
Note that the dot on a propagator in the diagram indicates that the propagator is doubled.

Derivatives of the above master integrals with respect to the squared mass arguments are useful. For the one-loop integrals and the two-loop vacuum integral:
\beq
\frac{\partial}{\partial x} {\bf A}(x) &=& (1 - \epsilon) {\bf A}(x)/x,
\label{eq:dboldAdx}
\\
\frac{\partial}{\partial x} {\bf B}(x,y) &=&
\frac{1}{\Delta_{sxy}} \left [
(1 - 2 \epsilon) (x-y-s) {\bf B}(x,y)
+ (1 - \epsilon) \left \{ (x+y-s) {\bf A}(x)/x - 2 {\bf A}(y) \right \}
\right ],\phantom{xxx}
\label{eq:dboldBdx}
\\
\frac{\partial}{\partial x} {\bf I}(x,y,z) &=&
\frac{1}{\Delta_{xyz}} \Bigl [
(1 - 2 \epsilon) (x - y - z) {\bf I}(x,y,z)
+ (1 - \epsilon) \bigl \{
(x - y + z) {\bf A}(x) {\bf A}(y)/x 
\nonumber \\ &&
+ 
(x + y - z) {\bf A}(x) {\bf A}(z)/x 
-2 {\bf A}(y) {\bf A}(z)
\bigr \} 
\Bigr ]
,\label{eq:dboldIdx}
\eeq
where the triangle function is
\beq
\Delta_{xyz} &=& x^2 + y^2 + z^2 - 2 x y - 2 x z - 2 y z.
\eeq
For the two-loop self-energy integrals, the simplest derivative is
\beq
\frac{\partial}{\partial x} {\bf S}(x,y,z) &=& - {\bf T}(x,y,z),
\label{eq:defboldT}
\eeq
since it is merely a definition.
The other derivatives of 2-loop integrals with respect to squared mass arguments are somewhat more complicated, and so the complete set of squared mass derivatives of 
${\bf A}, {\bf B}, {\bf I}, {\bf S}, {\bf T}, {\bf U},$ and ${\bf M}$ 
are provided in electronic form in an ancillary file {\tt derivs2loopbold}, for generic values of the squared masses. Also provided in that file are the derivatives with respect to $s$, which can be obtained from the squared mass derivatives by dimensional analysis.

Following the protocols given in the Introduction, the renormalized one-loop master integrals are now defined by subtracting the UV divergent parts and taking the limit:
\beq
A(x) &=& \lim_{\epsilon \rightarrow 0} \left [ {\bf A}(x) + x/\epsilon \right ]
\>=\> x L_x - x,
\\
B(x,y) &=& \lim_{\epsilon \rightarrow 0} \left [ {\bf B}(x,y) - 1/\epsilon \right ]
\>=\> -\int_0^1 dt\>\lnbar[tx + (1-t)y - t (1-t)s].
\eeq
The first of these equations allows us to trade $A(x)$ for $L_x$ at will, while the second can be easily evaluated analytically.
For the two-loop three-propagator renormalized master integral, define [following the general form of eqs.~(\ref{eq:defIrenormalized})-(\ref{eq:defIcounterterms})]:
\beq
S(x,y,z) &=& \lim_{\epsilon \rightarrow 0} \left [ {\bf S}(x,y,z) 
- S^{1,\,\rm div}(x,y,z) 
- S^{2,\,\rm div}(x,y,z) \right ],
\eeq
with contributions from one-loop and two-loop UV sub-divergences:
\beq
S^{1,\,\rm div}(x,y,z) &=& 
\frac{1}{\epsilon} \left [ {\bf A}(x) + {\bf A}(y) + {\bf A}(z)\right ],
\\
S^{2,\,\rm div}(x,y,z) &=& \frac{1}{2\epsilon^2} (x+y+z) + \frac{1}{2\epsilon}
(s/2 - x - y- z).
\eeq
The renormalized integrals $I(x,y,z)$ and $T(x,y,z)$ follow immediately from the above, as
\beq
I(x,y,z) &=& S(x,y,z) \Bigl |_{s=0},
\\
T(x,y,z) &=& -\frac{\partial}{\partial x} S(x,y,z).
\eeq
Next, define for the four-propagator renormalized integral:
\beq
U(w,x,y,z) &=& \lim_{\epsilon \rightarrow 0} \left [ {\bf U}(w,x,y,z) 
- U^{1,\,\rm div}(w,x,y,z) 
- U^{2,\,\rm div}(w,x,y,z) \right ],
\eeq
where the one-loop and two-loop UV sub-divergence contributions are 
\beq
U^{1,\,\rm div}(w,x,y,z) &=& \frac{1}{\epsilon} {\bf B}(w,x),
\\
U^{2,\,\rm div}(w,x,y,z) &=& -\frac{1}{2\epsilon^2} + \frac{1}{2\epsilon}.
\eeq
Finally, the five-propagator two-loop self-energy master integral is free of UV 
sub-divergences, so
\beq
M(v,w,x,y,z) &=& \lim_{\epsilon \rightarrow 0} {\bf M}(v,w,x,y,z).
\eeq

The derivatives of the renormalized integrals $A,B,I,S,T,U,M$ with
respect to each of their squared mass arguments, and $s$, can all be found in ref.~\cite{Martin:2003qz}. For convenience, they are also all provided in an ancillary file {\tt derivs2looprenorm} of the present paper. Also, the implicit dependences of the renormalized integrals on $Q$ are given by:
\beq
Q^2 \frac{\partial}{\partial Q^2} A(x) &=& -x,
\\
Q^2 \frac{\partial}{\partial Q^2} B(x,y) &=& 1,
\\
Q^2 \frac{\partial}{\partial Q^2} I(x,y,z) &=& A(x) + A(y) + A(z) - x - y - z,
\\
Q^2 \frac{\partial}{\partial Q^2} S(x,y,z) &=& A(x) + A(y) + A(z) - x - y - z + s/2,
\\
Q^2 \frac{\partial}{\partial Q^2} T(x,y,z) &=& -A(x)/x,
\\
Q^2 \frac{\partial}{\partial Q^2} U(w,x,y,z) &=& 1 + B(w,x),
\\
Q^2 \frac{\partial}{\partial Q^2} M(v,w,x,y,z) &=& 0.  
\eeq

It is also often convenient to define
\beq
V(w,x,y,z) &=& -\frac{\partial}{\partial x} U(w,x,y,z).
\eeq
Strictly speaking, this is not a master integral unless one of the squared masses vanishes, as it can be expressed in terms of the others (see the ancillary file {\tt derivs2looprenorm} of the present paper, or eqs.~(3.22)-(3.28) of ref.~\cite{Martin:2003qz}, for the explicit form). However, using it often simplifies expressions in practice. Moreover, if one of the squared masses vanishes, then $T(0,x,y)$ has a doubled massless propagator and is therefore IR divergent, so it is not available as an $\epsilon$-finite master integral. However, it can be replaced by either of the $\epsilon$-finite integrals $V(x,y,0,y)$ or $V(y,x,0,x)$,
or by the integral defined by the finite limit in which the mass-regulated IR divergence is subtracted:
\beq
\overline T(0,x,y) &=& \lim_{z \rightarrow 0} \left [ T(z,x,y) + L_z B(x,y) \right]. 
\eeq
The relation between $\overline T(0,x,y)$ and $V(x,y,0,y)$ is 
given by 
(see the Appendix of ref.~\cite{Martin:2003it}, which contains some similar identities):
\beq
\overline T(0,x,y) &=& 2 y \left [V(x,y,0,y) + (2 - L_y) \frac{\partial}{\partial y} B(x,y) \right ]
+ T(y,0,x) + L_y B(x,y) .
\eeq
For the special case $y=x$ in section \ref{sec:even} of the present paper, I will choose to use $V(x,x,0,x)$ as one of the master integrals. The program {\tt TSIL} \cite{Martin:2005qm} can be used for fast and accurate numerical evaluation of the renormalized $\epsilon$-finite master integrals $A,B,I,S,T,\overline{T}, U,V,M$ for any desired values of the arguments.

There are already several calculations that show by explicit example that the renormalized $\epsilon$-finite master integrals are the only ones needed to express renormalized two-loop self-energy observables in a general theory. These include the self-energies of scalars  refs.~\cite{Martin:2003it,Martin:2004kr,Martin:2005eg,Martin:2014cxa}, fermions in refs.~\cite{Martin:2005ch,Martin:2016xsp} and the Standard Model $W$ and $Z$ vector bosons in refs.~\cite{Martin:2015lxa,Martin:2015rea}. One might perhaps have thought that the integrals $A_\epsilon(x)$ and $B_{\epsilon}(x,y)$ defined by
\beq
{\bf A}(x) &=& -\frac{x}{\epsilon} + A(x) + \epsilon A_\epsilon(x) + \epsilon^2 A_{\epsilon^2}(x) \ldots
,
\\
{\bf B}(x,y) &=& \frac{1}{\epsilon} + B(x,y) + \epsilon B_\epsilon(x,y) + \epsilon^2 B_{\epsilon^2}(x,y) + \ldots
\eeq
might be necessary. However, this is not the case. In fact, it proved a useful check on the calculations listed above to observe the complete cancellation of $A_\epsilon$ and $B_{\epsilon}$ in the expressions for renormalized quantities. In a similar way, I have checked explicitly in ref.~\cite{SPMtoappear} that in the three-loop calculation of scalar and vector boson self-energy functions, one does not need $A_{\epsilon^2}$ or $B_{\epsilon^2}$, or the coefficients of positive powers of $\epsilon$ in the two-loop functions ${\bf I}, {\bf S}, {\bf T}, {\bf U}, {\bf M}$, either.\footnote{In the particular case of ${\bf A}(x)$ and ${\bf B}(x,y)$, the expansions to all orders in $\epsilon$ are known; see ref.~\cite{Davydychev:2000na} for the latter in terms of Nielsen polylogarithms. The point being made here is that these should never be needed.} All occurrences of them cancel. Among the one-loop and two-loop integral functions, only $A,B,I,S,T,U,M$ (and either $\overline T$ or $V$ if a squared mass vanishes) are needed for the self-energy expressed in terms of renormalized couplings and masses, even at three-loop order. A similar statement holds for the general three-loop effective potential, as shown explicitly in ref.~\cite{Martin:2017lqn}. This is presumably true at all orders in perturbation theory. It should also hold in on-shell and hybrid type renormalization schemes, since they can be related to the modified minimal subtraction scheme by redefinitions involving renormalized physical quantities. 

\subsection{Three-loop self-energy integrals\label{subsec:threeSE}}

\begin{figure}[!p]
\begin{center}
\includegraphics[width=17cm,angle=0]{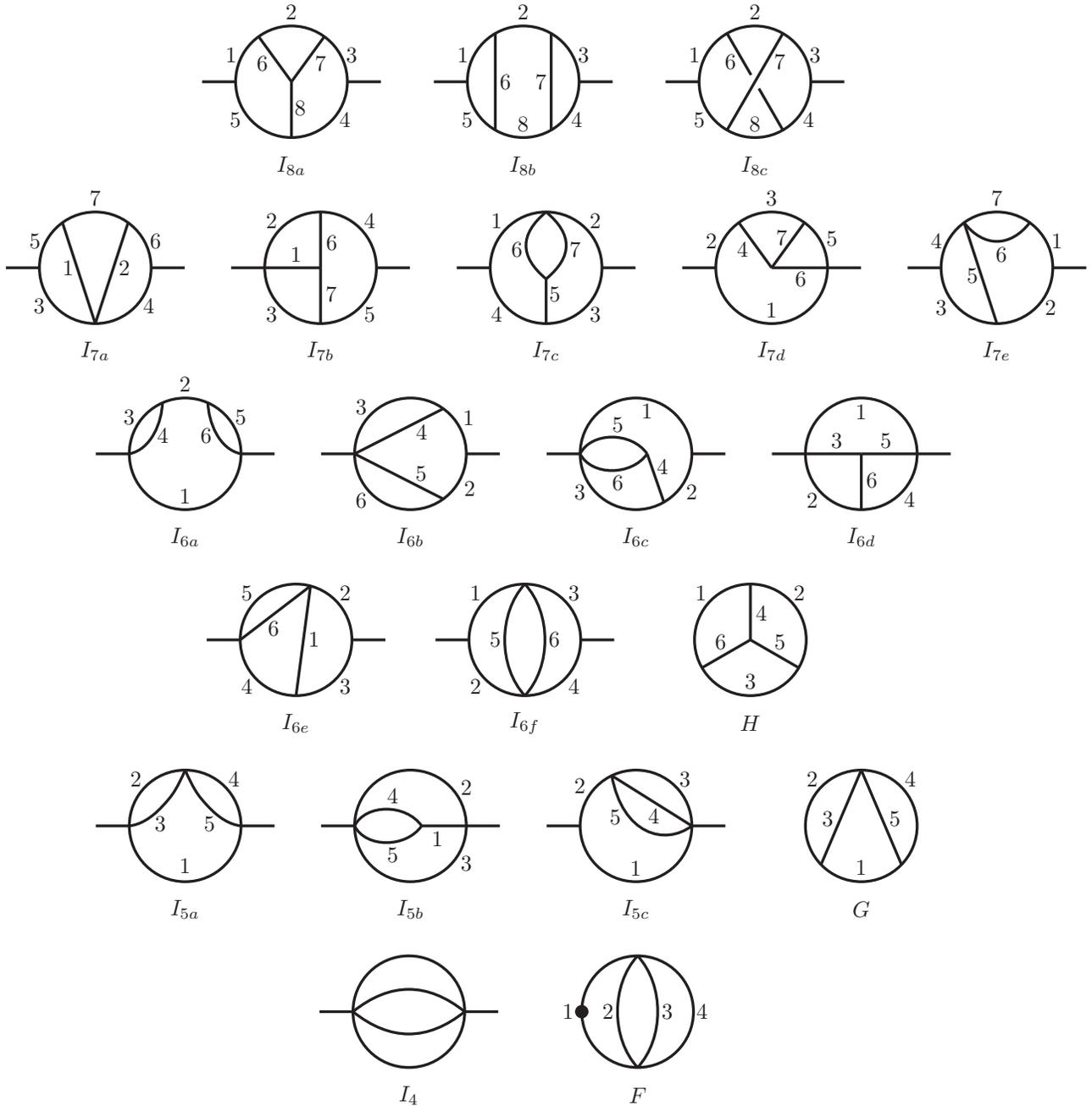}
\end{center}
\caption{\label{fig:threelooptopologies}
Topologies for three-loop self-energy and vacuum scalar integrals. The integer labels on the internal lines denote the ordering of propagator squared mass arguments adopted in this paper. The vacuum integrals $F$, $G$, and $H$ follow the conventions also used in refs.~\cite{Martin:2016bgz} and \cite{Martin:2017lqn}.
For the self-energy integrals with names containing $I$, the first (integer) subscript in the name is the number of internal propagator lines. Not shown are topologies that factorize into products of 1-loop and 2-loop integrals. }
\end{figure}

Consider scalar self-energy functions at three-loop order, which can have the topologies in Fig.~\ref{fig:threelooptopologies}, with denominators arising from arbitrary powers of the propagators shown, and numerators that are polynomials in scalar products of the external 4-momenta $p^\mu$, and the loop integration momenta $q^\mu$, $k^\mu$, $r^\mu$. In this paper, I define ``candidate master" 3-loop scalar self-energy integrals as follows.

First, we have three 8-propagator scalar integrals 
\beq
&&{\bf I}_{8a}(x_1,x_2,x_3,x_4,x_5,x_6,x_7,x_8) 
\>=\>
\int_q \int_k \int_r \,
\Bigl \{
[q^2 + x_1] [k^2 + x_2] [r^2 + x_3] 
[(r-p)^2 + x_4] 
\nonumber \\ &&
\qquad
[(q-p)^2 + x_5] 
[(q-k)^2 + x_6]
[(k-r)^2 + x_7] [(r-q)^2 + x_8] \Bigr \}^{-1}
\label{eq:bfI8a}
\\
&&{\bf I}_{8b}(x_1,x_2,x_3,x_4,x_5,x_6,x_7,x_8) 
\>=\>
\int_q \int_k \int_r \,
\Bigl \{
[q^2 + x_1] [k^2 + x_2] [r^2 + x_3] 
[(r-p)^2 + x_4] 
\nonumber \\ &&
\qquad
[(q-p)^2 + x_5] 
[(q-k)^2 + x_6]
[(k-r)^2 + x_7] 
[(k-p)^2 + x_8] \Bigr \}^{-1}
\label{eq:bfI8b}
\\
&&
{\bf I}_{8c}(x_1,x_2,x_3,x_4,x_5,x_6,x_7,x_8) 
\>=\>
\int_q \int_k \int_r \,
\Bigl \{
[q^2 + x_1] [k^2 + x_2] [r^2 + x_3] 
[(r-p)^2 + x_4] 
\nonumber \\ &&
\qquad
[(q-p)^2 + x_5] 
[(q-k)^2 + x_6]
[(k-r)^2 + x_7] [(q+r-k-p)^2 + x_8] \Bigr \}^{-1}
,
\label{eq:bfI8c}
\eeq
as depicted in the top row of Figure
\ref{fig:threelooptopologies}. The last of these has a non-planar topology.

Besides the 8-propagator integrals, it is necessary to also include all integrals obtained from them by 
removing one of more of the scalar propagator factors, as shown in the remaining rows of Figure
\ref{fig:threelooptopologies}. In particular, this figure defines the 7-propagator integrals labeled ${\bf I}_{7a}$, ${\bf I}_{7b}$, ${\bf I}_{7c}$, ${\bf I}_{7d}$, and ${\bf I}_{7e}$, the 6-propagator integrals shown in the third and fourth rows, the 5-propagator integrals in the fifth row, and the 4-propagator integrals in the sixth row. In each case, the ordering of the squared mass arguments $x_1, x_2,\ldots$, is indicated by the integer labels. The external momentum invariant $s$ and the renormalization scale $Q$ are the same in each case, so they are not included explicitly in the list of arguments. 

However, the integrals just defined (with unit numerator, and denominators with only
single powers of propagators) are not sufficient. 
In addition, the list of candidate master integrals
includes all integrals obtained from the ones shown by doubling one of the propagators, which is the same as taking the negative of the derivative with respect to the corresponding squared mass argument. This is indicated by adding the corresponding integer to the end of the subscript in the integral name, for example:
\beq
{\bf I}_{41}(w,x,y,z) &=& -\frac{\partial}{\partial w} {\bf I}_{4}(w,x,y,z),
\label{eq:defboldI41}
\\
{\bf I}_{5a1}(v,w,x,y,z) &=& -\frac{\partial}{\partial v} {\bf I}_{5a}(v,w,x,y,z),
\\
{\bf I}_{6c5}(u,v,w,x,y,z) &=& -\frac{\partial}{\partial y} {\bf I}_{6c}(u,v,w,x,y,z).
\eeq
However, for the eight-propagator integrals, I find that derivatives of ${\bf I}_{8a}$ and ${\bf I}_{8b}$ are never necessary, and only one of the two distinct derivatives of ${\bf I}_{8c}$ is necessary, which can be chosen to be ${\bf I}_{8c1}$. Besides the preceding, a few other integrals are useful in the general case. For the four-propagator topology only, define an integral with both of the first two propagators doubled:
\beq
{\bf I}_{412}(w,x,y,z) &=& \frac{\partial^2}{\partial w\partial x} {\bf I}_{4}(w,x,y,z),
\label{eq:defboldI412}
\eeq
and one with the first propagator tripled:
\beq
{\bf I}_{411}(w,x,y,z) &=& \frac{1}{2} \frac{\partial^2}{\partial w^2} {\bf I}_{4}(w,x,y,z)
.
\label{eq:defboldI411}
\eeq
Finally, for the non-planar 8-propagator topology (8c) only, it is evidently necessary 
(see, for example, the case of all masses equal, considered in section \ref{sec:allmassive} below) to define a master integral with $p\cdot k$ in the integrand numerator:
\beq
&&
{\bf I}_{8c}^{pk}(x_1,x_2,x_3,x_4,x_5,x_6,x_7,x_8)
\>=\>
\int_q \int_k \int_r \,
p\cdot k \>
\Bigl \{
[q^2 + x_1] [k^2 + x_2] [r^2 + x_3] 
[(r-p)^2 + x_4] 
\nonumber \\ &&
\qquad
[(q-p)^2 + x_5] 
[(q-k)^2 + x_6]
[(k-r)^2 + x_7] [(q+r-k-p)^2 + x_8] \Bigr \}^{-1}
.
\eeq
Together with products of 1-loop and 2-loop integrals, this concludes\footnote{Integrals that would be redundant by symmetry are not included; for example, there is no ${\bf I}_{5a3}$, because ${\bf I}_{5a3}(v,w,x,y,z)$ would be the same as ${\bf I}_{5a2}(v,x,w,y,z)$.} the listing of the 3-loop candidate master self-energy integrals. 
Not all of these will be linearly independent, so the number of actual master integrals will always be smaller, depending on the choice of squared mass arguments. Also, integrals with IR divergences cannot be $\epsilon$-finite masters. The IR divergent
cases include any integral with a doubled massless propagator, and also integrals ${\bf I}_{7d}(x_1,0,0,0,0,0,0)$ for any $x_1$, and ${\bf I}_{8a}(0,0,0,x_4,x_5,0,0,0)$ for any $x_4$ and $x_5$. The choice of the masters from among the candidate masters is not unique. Furthermore, the possibility of identities that could eliminate one or more of the putative masters is not always easy to rule out.
However, it does no harm (except, in some cases, some avoidable complication) to include extra masters beyond a minimal set. In some cases, including extra masters may lead to more compact expressions.

I now proceed to define the renormalized $\epsilon$-finite masters. First, for the vacuum integrals $F(w,x,y,z)$, $G(v,w,x,y,z)$, and $H(u,v,w,x,y,z)$, the definitions have already been provided in section II of ref.~\cite{Martin:2016bgz}, and also coincide with the definitions given below for $I_{41}(w,x,y,z)$, $I_{5a}(v,w,x,y,z)$, and $I_{6d}(u,v,x,w,z,y)$ with $s=0$. The program {\tt 3VIL} provides for the fast and accurate evaluation of the functions $F$, $G$, and $H$ with arbitrary arguments, including various special cases given originally in refs.~\cite{Broadhurst:1991fi,Avdeev:1994db,Fleischer:1994dc,Avdeev:1995eu,Broadhurst:1998rz,Fleischer:1999mp,Steinhauser:2000ry,Schroder:2005va,Davydychev:2003mv,Kalmykov:2005hb,Kalmykov:2006pu,Bytev:2009mn,Bekavac:2009gz,Bytev:2009kb,Bytev:2011ks,Grigo:2012ji}.  For another approach to the numerical calculation of three-loop vacuum and self-energy integrals with general masses, see refs.~\cite{Bauberger:2017nct} and \cite{Bauberger:2019heh} respectively.

For the four-propagator self-energy integral [compare to the general form of eqs.~(\ref{eq:defIrenormalized})-(\ref{eq:defIcounterterms})]:
\beq
I_4(w,x,y,z) &=& \lim_{\epsilon \rightarrow 0} \Bigl [
{\bf I}_4(w,x,y,z) 
- {\bf I}_4^{1,\rm div}(w,x,y,z)
- {\bf I}_4^{2,\rm div}(w,x,y,z)
- {\bf I}_4^{3,\rm div}(w,x,y,z)
\Bigr ]
,
\phantom{xxxx}
\eeq
where the 1-loop, 2-loop, and 3-loop UV sub-divergence subtractions are:
\beq
{\bf I}_4^{1,\rm div}(w,x,y,z) &=& 
\frac{1}{\epsilon} 
\Bigl [
{\bf A}(w) {\bf A}(x) 
+
{\bf A}(w) {\bf A}(y)
+{\bf A}(w) {\bf A}(z)
+{\bf A}(x) {\bf A}(y)
+{\bf A}(x) {\bf A}(z)
\nonumber \\ &&
+{\bf A}(y) {\bf A}(z)
\Bigr ]
,
\\
{\bf I}_4^{2,\rm div}(w,x,y,z) &=&
\left \{ \left (\frac{1}{2\epsilon^2} - \frac{1}{2\epsilon} \right )(x+y+z) 
+ \frac{1}{4 \epsilon} (s+w) \right \}
{\bf A}(w) 
+   \mbox{(three permutations)}, \phantom{xxxx}
\\
{\bf I}_4^{3,\rm div}(w,x,y,z) &=&
\frac{s^2}{36\epsilon} + \left ( \frac{1}{6 \epsilon^2} - \frac{1}{8\epsilon} \right ) s (w+x+y+z)
+ \left (\frac{1}{6 \epsilon^2} - \frac{3}{8 \epsilon} \right )
(w^2 + x^2 + y^2 + z^2)
\nonumber \\ && 
+ \left ( \frac{1}{3\epsilon^3} - \frac{2}{3\epsilon^2} + \frac{1}{3\epsilon} \right )
(w x + w y + w z + x y + x z + y z)
.
\eeq
The expressions for the renormalized integrals $I_{41}$, $I_{411}$, and $I_{412}$ are easily obtained from the above by taking derivatives with respect to the squared mass arguments, following from eqs.~(\ref{eq:defboldI41}), (\ref{eq:defboldI412}) and (\ref{eq:defboldI411}), by making use of eq.~(\ref{eq:dboldAdx}).

The remaining renormalized masters are constructed in an entirely analogous way.
For the five-propagator integrals, the subtractions before taking the limit $\epsilon \rightarrow 0$ are:
\beq
{\bf I}_{5a}^{1,\rm div}(v,w,x,y,z) &=&
\frac{1}{\epsilon} \bigl [{\bf S}(v,w,x) + {\bf S}(v,y,z) \bigr ]
,
\\
{\bf I}_{5a}^{2,\rm div}(v,w,x,y,z) &=&
-\frac{1}{\epsilon^2} {\bf A}(v)
+\left (\frac{1}{2 \epsilon} - \frac{1}{2 \epsilon^2}\right ) 
\bigl  [{\bf A}(w) + {\bf A}(x)+ {\bf A}(y) + {\bf A}(z)  \bigr ]
,
\\
{\bf I}_{5a}^{3,\rm div}(v,w,x,y,z) &=&
\left ( -\frac{1}{6 \epsilon^2} + \frac{1}{12 \epsilon} \right ) s
+
\left ( -\frac{1}{6 \epsilon^3} + \frac{1}{2 \epsilon^2} - \frac{2}{3\epsilon} \right )
(w+x+y+z)
\nonumber \\ && 
+
\left ( -\frac{1}{3 \epsilon^3} + \frac{1}{3 \epsilon^2} + \frac{1}{3\epsilon} \right ) v
,
\eeq
and
\beq
{\bf I}_{5b}^{1,\rm div}(v,w,x,y,z) &=&
\frac{1}{\epsilon} \bigl [{\bf S}(v,w,x) + {\bf I}(v,y,z) \bigr ]
,
\\
{\bf I}_{5b}^{2,\rm div}(v,w,x,y,z) &=&
-\frac{1}{\epsilon^2} {\bf A}(v) 
+\left (\frac{1}{2 \epsilon} - \frac{1}{2 \epsilon^2}\right ) 
\bigl [{\bf A}(w) + {\bf A}(x)+ {\bf A}(y) + {\bf A}(z)  \bigr ]
,
\\
{\bf I}_{5b}^{3,\rm div}(v,w,x,y,z) &=&
\left ( -\frac{1}{12 \epsilon^2} + \frac{5}{24 \epsilon} \right ) s
+
\left ( -\frac{1}{6 \epsilon^3} + \frac{1}{2 \epsilon^2} - \frac{2}{3\epsilon} \right )
(w+x+y+z)
\nonumber \\ && 
+
\left ( -\frac{1}{3 \epsilon^3} + \frac{1}{3 \epsilon^2} + \frac{1}{3\epsilon} \right ) v
,
\eeq
and
\beq
{\bf I}_{5c}^{1,\rm div}(v,w,x,y,z) &=&
\frac{1}{\epsilon} {\bf B}(v,w) \bigl [{\bf A}(x) + {\bf A}(y) + {\bf A}(z) \bigr ] 
,
\\
{\bf I}_{5c}^{2,\rm div}(v,w,x,y,z) &=&
-\frac{1}{4\epsilon} {\bf A}(v) 
+ \left (\frac{1}{2 \epsilon} - \frac{1}{2 \epsilon^2}\right ) 
  \bigl [{\bf A}(x)+ {\bf A}(y) + {\bf A}(z)  \bigr ]
\nonumber \\ &&
+ \left [\left (\frac{1}{2\epsilon^2} - \frac{1}{2\epsilon} \right ) (x+y+z) + 
\frac{1}{4\epsilon} w \right] {\bf B}(v,w)  
,
\\
{\bf I}_{5c}^{3,\rm div}(v,w,x,y,z) &=&
-\frac{1}{12 \epsilon} s + 
\left (-\frac{1}{6 \epsilon^2} + \frac{3}{8\epsilon}\right ) (v+w)
+
\left ( -\frac{1}{3 \epsilon^3} + \frac{2}{3 \epsilon^2} - \frac{1}{3\epsilon} \right )
(x+y+z)
.
\phantom{xxx}
\eeq
Again the corresponding expressions for $I_{5a1}$, $I_{5a2}, \ldots$ are obtained
from the above by taking derivatives with respect to the appropriate squared mass arguments in the obvious way, making use of eqs.~(\ref{eq:dboldAdx})-(\ref{eq:dboldIdx}) and (\ref{eq:defboldT}).

The subtractions for the six-propagator three-loop self-energy integrals are given by
\beq
{\bf I}_{6a}^{1,\rm div}(u,v,w,x,y,z) &=&
\frac{1}{\epsilon} \bigl [ {\bf U}(u,v,w,x) +  {\bf U}(u,v,y,z)\bigr ] 
,
\\
{\bf I}_{6a}^{2,\rm div}(u,v,w,x,y,z) &=&
-\frac{1}{\epsilon^2} {\bf B}(u,v)
,
\\
{\bf I}_{6a}^{3,\rm div}(u,v,w,x,y,z) &=&
\frac{1}{3 \epsilon^3} - \frac{1}{3 \epsilon^2} - \frac{1}{3\epsilon} 
,
\eeq
and
\beq
{\bf I}_{6b}^{1,\rm div}(u,v,w,x,y,z) &=&
\frac{1}{\epsilon} \bigl [ {\bf U}(u,v,y,z) + {\bf U}(v,u,w,x) \bigr ] 
,
\\
{\bf I}_{6b}^{2,\rm div}(u,v,w,x,y,z) &=&
-\frac{1}{\epsilon^2} {\bf B}(u,v)
,
\\
{\bf I}_{6b}^{3,\rm div}(u,v,w,x,y,z) &=&
\frac{1}{3 \epsilon^3} - \frac{1}{3 \epsilon^2} - \frac{1}{3\epsilon} 
,
\eeq
and
\beq
{\bf I}_{6c}^{1,\rm div}(u,v,w,x,y,z) &=&
\frac{1}{\epsilon} {\bf U}(u,v,w,x)
,
\\
{\bf I}_{6c}^{2,\rm div}(u,v,w,x,y,z) &=&
\left (\frac{1}{2\epsilon} - \frac{1}{2\epsilon^2} \right ){\bf B}(u,v)
,
\\
{\bf I}_{6c}^{3,\rm div}(u,v,w,x,y,z) &=&
\frac{1}{6 \epsilon^3} - \frac{1}{2 \epsilon^2} + \frac{2}{3\epsilon},
\eeq
and
\beq
{\bf I}_{6d}^{1,\rm div}(u,v,w,x,y,z) &=&
{\bf I}_{6d}^{2,\rm div}(u,v,w,x,y,z) \>=\>
0,
\\
{\bf I}_{6d}^{3,\rm div}(u,v,w,x,y,z) &=&
{2 \zeta_3}/{\epsilon} ,
\eeq
and
\beq
{\bf I}_{6e}^{1,\rm div}(u,v,w,x,y,z) &=&
\frac{1}{\epsilon} {\bf U}(v,w,u,x)
,
\\
{\bf I}_{6e}^{2,\rm div}(u,v,w,x,y,z) &=&
\left (\frac{1}{2\epsilon} - \frac{1}{2\epsilon^2} \right ){\bf B}(v,w)
,
\\
{\bf I}_{6e}^{3,\rm div}(u,v,w,x,y,z) &=&
\frac{1}{6 \epsilon^3} - \frac{1}{2 \epsilon^2} + \frac{2}{3\epsilon}
,
\eeq
and
\beq
{\bf I}_{6f}^{1,\rm div}(u,v,w,x,y,z) &=&
\frac{1}{\epsilon} {\bf B}(u,v){\bf B}(w,x)
,
\\
{\bf I}_{6f}^{2,\rm div}(u,v,w,x,y,z) &=&
\left (\frac{1}{2\epsilon} - \frac{1}{2\epsilon^2} \right )
\bigl [ {\bf B}(u,v) + {\bf B}(w,x) \bigr ]
,
\\
{\bf I}_{6f}^{3,\rm div}(u,v,w,x,y,z) &=&
\frac{1}{3 \epsilon^3} - \frac{2}{3 \epsilon^2} + \frac{1}{3\epsilon}
.
\eeq
Again the corresponding expressions for $I_{6a1}$, $I_{6a3}, \ldots$ are obtained
from the above by taking derivatives with respect to the appropriate squared mass arguments, making use of eq.~(\ref{eq:dboldBdx})
and the corresponding formulas for the derivatives of ${\bf U}$, which can be found
in the ancillary file {\tt derivs2loopbold}.

For the seven-propagator three-loop integrals, only ${\bf I}_{7c}$ and ${\bf I}_{7e}$
have UV divergences. The corresponding subtractions are:
\beq
{\bf I}_{7c}^{1,\rm div}(t,u,v,w,x,y,z) &=&
\frac{1}{\epsilon} {\bf M}(t,u,w,v,x)
,
\\
{\bf I}_{7c}^{2,\rm div}(t,u,v,w,x,y,z) &=&
{\bf I}_{7c}^{3,\rm div}(t,u,v,w,x,y,z) \>=\> 0
,
\eeq
and
\beq
{\bf I}_{7e}^{1,\rm div}(t,u,v,w,x,y,z) &=&
\frac{1}{\epsilon} {\bf M}(t,w,u,v,x)
,
\\
{\bf I}_{7e}^{2,\rm div}(t,u,v,w,x,y,z) &=&
{\bf I}_{7e}^{3,\rm div}(t,u,v,w,x,y,z) \>=\> 0
.
\eeq
The corresponding expressions for $I_{7c1}$, $I_{7c3}, \ldots$ are obtained by making use of the formulas for the derivatives of ${\bf M}$ with respect to its squared mass arguments, as given in the ancillary file {\tt derivs2loopbold}.

There are no UV divergences, and therefore no subtractions, for the seven-propagator candidate masters ${\bf I}_{7a}$, ${\bf I}_{7b}$, ${\bf I}_{7d}$, and the
eight-propagator integrals ${\bf I}_{8a}$, ${\bf I}_{8b}$, ${\bf I}_{8c}$, and ${\bf I}_{8c}^{pk}$. The renormalized $\epsilon$-finite candidate masters 
$I_{7a}$, $I_{7b}$, $I_{7d}$, $I_{8a}$, $I_{8b}$, $I_{8c}$, and $I_{8c}^{pk}$
are therefore just the $\epsilon\rightarrow 0$ limits of the bold-faced integrals.
The same holds for arbitrary derivatives of them with respect to their squared mass arguments.

The renormalized masters that required UV subtractions depend on the scale $Q$, although this dependence is suppressed from the list of arguments. The results for $\displaystyle Q \frac{\partial}{\partial Q}$ are determined by the above definitions, and are given in an ancillary file {\tt QddQ}
provided with this paper.

One approach is to treat all squared masses as completely generic, in which case IR divergences are regularized by the non-zero values assigned to gauge bosons and chiral fermions, which can be sent to zero at the end of calculation. If, on the other hand, we impose special relations among the masses (typically, that some of them vanish, and/or that others are equal to each other) then expressions can be much simpler but it is not completely trivial to choose an $\epsilon$-finite basis for the master integrals. We will do this in some notable special cases in sections \ref{sec:zeromass}, \ref{sec:allmassive}, \ref{sec:odd}, and \ref{sec:even}.

To conclude this section, note that the relationship between the original (bold-faced) and renormalized integrals can of course be inverted, in an expansion in $\epsilon$. For example, for the four-propagator self-energy integral, one can write:
\beq
{\bf I}_4(w,x,y,z) &=& 
\frac{1}{\epsilon^3} I_4^{(-3)}(w,x,y,z) + 
\frac{1}{\epsilon^2} I_4^{(-2)}(w,x,y,z) +
\frac{1}{\epsilon} I_4^{(-1)}(w,x,y,z) +
I_4^{(0)} (w,x,y,z) + \ldots
\nonumber \\ &&
\eeq
where
\beq
I_4^{(-3)}(w,x,y,z) &=&
\left (w x +  w y + w z  + x y   + x z  + y z \right )/3
,
\\
I_4^{(-2)}(w,x,y,z) &=&
- s (w + x + y + z)/12 
- (w^2 + x^2 + y^2 + z^2)/12
\nonumber \\ &&
+ (w x +  w y  + x y  + w z  + x z  + y z)/3
- \Bigl [(x+y+z) A(w)
\nonumber \\ &&
+ (w+y+z) A(x)
+ (w+x+z) A(y)
+ (w +x+y) A(z)\Bigr ]/2
,
\\
I_4^{(-1)}(w,x,y,z) 
&=&
{s^2}/{36} - s(w+x+y+z)/8 
-3 (w^2 + x^2 + y^2 + z^2)/8 
\nonumber \\ && 
+ (w x +  w y + w z  + x y   + x z  + y z )/3
+ (s + w - 2x - 2y- 2z) A(w)/4
\nonumber \\ && 
+ (s + x - 2w - 2y- 2z) A(x)/4
+ (s + y - 2w - 2x- 2z) A(y)/4
\nonumber \\ &&
+ (s + z - 2w - 2x- 2y) A(z)/4
+ A(w) A(x)
+ A(w) A(y)
+ A(w) A(z)
\nonumber \\ &&
+ A(x) A(y)
+ A(x) A(z)
+ A(y) A(z)
- (x+y+z) \Aeps(w)/2
\nonumber \\ &&
- (w+y+z) \Aeps(x)/2
- (w+x+z) \Aeps(y)/2
- (w+x+y) \Aeps(z)/2
,
\\
I_4^{(0)} (w,x,y,z)
&=&
I_4(w,x,y,z)
+ s \bigl [ \Aeps(w) + \Aeps(x) + \Aeps(y) + \Aeps(z) \bigr ]/4
\nonumber \\ &&
+ A(w) \bigl[\Aeps(x) + \Aeps(y) + \Aeps(z)\bigr]
+ A(x) \bigl[\Aeps(w) + \Aeps(y) + \Aeps(z)\bigr]
\nonumber \\ &&
+ A(y) \bigl[\Aeps(w) + \Aeps(x) + \Aeps(z)\bigr]
+ A(z) \bigl[\Aeps(w) + \Aeps(x) + \Aeps(y)\bigr]
\nonumber \\ &&
- (x+y+z) \bigl[\Aeps(w) + \Aepstwo(w)\bigr]/2
- (w+y+z) \bigl[\Aeps(x) + \Aepstwo(x)\bigr]/2
\nonumber \\ &&
- (w+x+z) \bigl[\Aeps(y) + \Aepstwo(y)\bigr]/2
- (w+x+y) \bigl[\Aeps(z) + \Aepstwo(z)\bigr]/2
\nonumber \\ &&
+  [w \Aeps(w) + x \Aeps(x) + y \Aeps(y) + z \Aeps(z)]/4.
\label{eq:I40expansion}
\eeq
However, I emphasize that integral functions like $I_4^{(0)}$, which occur as the coefficient of $\epsilon^0$ in the expansion of the original integrals, are quite sub-optimal for expressing results for renormalized physical quantities. This is because
writing 3-loop results in terms of such integrals requires that the expressions will also include $A_{\epsilon}(x)$, $A_{\epsilon^2}(x)$, as can be seen from eq.~(\ref{eq:I40expansion}). Similarly, for quantities with five or more propagators, $B_\epsilon(x,y)$, $B_{\epsilon^2}(x,y)$, $S_{\epsilon}(x,y,z)$ etc.~will appear, which involve the coefficients of positive powers of $\epsilon$ from integrals at lower loop order than the calculation being performed. The big advantage of organizing results in terms of the renormalized $\epsilon$-finite integrals like $I_4(w,x,y,z)$ is that such coefficients of positive powers of $\epsilon$ are never needed. This property should persist to arbitrary loop order.

\subsection{Numerical evaluation by differential equations\label{subsec:numericaleval}}

For the case of self-energy renormalized $\epsilon$-finite master integrals,
the derivatives with respect to $s$ can be expressed as linear combinations of them:
\beq
\frac{d}{ds} I_j &=& \sum_k c_{jk} I_k,
\label{eq:dIjds}
\eeq
where the coefficients $c_{jk}$ are rational functions of $s$ and the internal
propagator squared masses. (Note that the $c_{jk}$
do not depend on $\epsilon$ in this approach, since the $I_j$ are independent of $\epsilon$ by construction.)
Once these coefficients have been found, as we will do below in various special cases, then one can solve the coupled first-order differential equations numerically, using a Runge-Kutta or similar algorithm. 

The initial boundary conditions for the numerical integration of the differential equations can be obtained at or near $s=0$. If $s=0$ is not a threshold for any of the master integrals under consideration, then the initial boundary conditions can typically be set at $s=0$ in terms of the vacuum integral masters, available in the notation of the present paper from ref.~\cite{Martin:2016bgz}, incorporating some original analytic calculations for special cases from refs.~\cite{Broadhurst:1991fi,Avdeev:1994db,Fleischer:1994dc,Avdeev:1995eu,Broadhurst:1998rz,Fleischer:1999mp,Steinhauser:2000ry,Schroder:2005va,Davydychev:2003mv,Kalmykov:2005hb,Kalmykov:2006pu,Bytev:2009mn,Bekavac:2009gz,Bytev:2009kb,Bytev:2011ks,Grigo:2012ji}. If $s=0$ is a threshold for one or more of the masters, then one can instead choose an initial boundary condition at some small $s_0$. To do so, the self-energy masters can be written as a series expansion in small $s$, with coefficients obtained using the same differential equations (\ref{eq:dIjds}) and expressed in terms of the vacuum integral masters.  The initial conditions are then evaluated at an appropriate $s=s_0$ within the radius of convergence of the series.

\begin{figure}[!t]
\begin{center}
\includegraphics[width=9.6cm,angle=0]{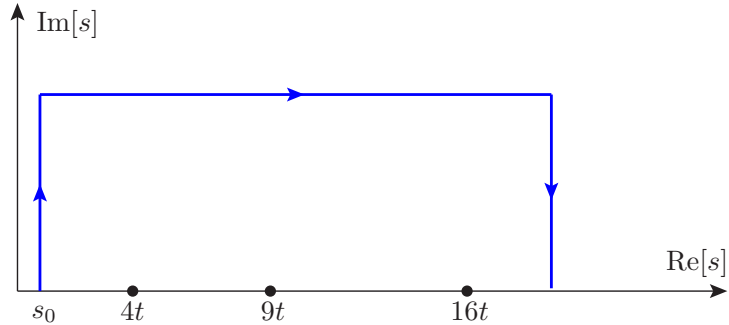}
\end{center}
\vspace{-0.3cm}
\begin{minipage}[]{0.97\linewidth}
\caption{\label{fig:contour} Path in the complex $s$ plane for numerical integration of first-order coupled differential equations for the master integrals. The integration
starts at $s=s_0$ chosen within the domain of convergence of the small-$s$ series expansion. The contour avoids singular points in the differential equations (shown here as occurring at $s=4t$, $9t$, and $16t$, as in section \ref{sec:allmassive}) by proceeding in the upper-half complex plane, giving the correct branch cut for an infinitesimal imaginary part of $s$ at the end of the path.}
\end{minipage}
\end{figure}

In order to obtain the correct imaginary parts of the masters, one can follow the strategy introduced in ref.~\cite{Caffo:2002ch,Caffo:2002wm} by using a contour in the upper-half complex plane for the Runge-Kutta integration, thus avoiding branch cuts and other special points on the Im$[s]=0$ line, as shown in Figure \ref{fig:contour}. 
This procedure is the one used by the program {\tt TSIL} \cite{Martin:2005qm},
to find the two-loop self-energy renormalized masters for general squared masses. 

I have constructed a similar (but not particularly
well-optimized) mathematica program to compute the three-loop master integrals
for the special cases considered in sections \ref{sec:allmassive}, \ref{sec:odd},
and \ref{sec:even} below. (It is left as an exercise for the reader to do the same.) In principle, this should be straightforward in more general cases, although the coefficients will be considerably more complicated and some optimization (including partial fraction decomposition of coefficient functions, and certain specialized Runge-Kutta routines designed to minimize numerical problems and improve calculation speed near thresholds) may be needed.

There are several advantages of the numerical evaluation method outlined above.
First, all of the masters descended from a given topology are obtained simultaneously as the result of a single calculation.
Second, the Runge-Kutta method (and refinements thereof) tends to be faster and more accurate than multi-dimensional integral methods.
Third, changing the contour in the upper-half complex $s$ plane allows for consistency checks and numerical error estimates.    

\section{The case of massless internal propagators\label{sec:zeromass}}
\setcounter{equation}{0}
\setcounter{figure}{0}
\setcounter{table}{0}
\setcounter{footnote}{1}

In this section, I review the case that all internal masses vanish, where the integrals are all known analytically. This will help to illustrate the connection between the general notation and the known results in this special case.

With all vanishing masses, the renormalized $\epsilon$-finite master integrals do not include doubled propagators, as these are IR divergent. The scalar integrals for the topologies $I_{7d}$ and $I_{8a}$
are also easily seen to be IR divergent, despite not having doubled massless propagators. By using the IBP relations and known results \cite{Chetyrkin:1980pr,Tkachov:1981wb,Chetyrkin:1981qh,Gorishnii:1989gt,Larin:1991fz,Baikov:2010hf}, one finds that there are only four independent renormalized $\epsilon$-finite master integrals. They can be chosen to be $B$, $M$, $I_{6d}$, and $I_{7a}$, although there are clearly other equally valid choices. In terms of them, the renormalized $\epsilon$-finite candidate master scalar integrals are (suppressing all internal squared mass arguments in this section, since they all vanish):
\beq
B &=& 2 - L_{-s}
,
\label{eq:Ballmassless}
\\
S &=& s \left (B/2 + 5/8\right )
,
\\
U &=& B^2/2 + B + 3/2,
,
\\
M &=& - 6 \zeta_3/s,
\\
I_{4} &=& s^2 \left (B/12 + 35/216 \right )
,
\\
I_{5a} &=& s \left (B^2/2 + 3 B/2 + 47/24 \right )
,
\\
I_{5b} &=& s \left ( B^2/4 + 5 B/4 + 103/48 \right )
,
\\
I_{5c} &=& -s \left (B/4 + 13/24 \right ) 
,
\\
I_{6a} &=& B^3/3 + B^2 + 2 B + 2\zeta_3/3 + 5/3
,
\\
I_{6b} &=& B^3/3 + B^2 + 2 B - 4\zeta_3/3 + 5/3
,
\\
I_{6c} &=& B^3/6 + B^2 + 7 B/2 - 2\zeta_3/3 + 14/3
,
\\
I_{6d} &=& 3 \zeta_4 + 6 \zeta_3 B
,
\\
I_{6e} &=& B^3/6 + B^2 + 7 B/2 - 14\zeta_3/3 + 14/3
,
\\
I_{6f} &=& B^3/3 + B^2 + B + 14\zeta_3/3 - 7/3
,
\\
I_{7a} &=& I_{7b} \>=\> -20 \zeta_5/s
,
\\
I_{7c} &=& I_{7e} \>=\> -6 \zeta_3 B/s
, 
\\
I_{8b} &=& I_{8c} \>=\> 20 \zeta_5/s^2,
\label{eq:I8b8czeromass}
\\
I_{8c}^{pk} &=& -5 \zeta_5/s.
\label{eq:I8cpkallmassless}
\eeq
Alternatively, the independent master integral quantities can be taken to be $L_{-s} \equiv \lnbar(s)- i\pi$ from 1-loop order, $\zeta_3$ at two-loop order, and $\zeta_4$ and $\zeta_5$ at three-loop order.

The integrals $I_{7d}$ and $I_{8a}$ can also be evaluated with the results
\beq
s I_{7d} &=& -s^2 I_{8a} \>=\>
3 \zeta_4 + \zeta_3 \left (\frac{2}{\epsilon_{\rm IR}}  + 6 B - 12 \right )
.
\qquad\phantom{x}
\eeq
Here the $1/\epsilon_{\rm IR}$ poles remain uncanceled (there are no UV sub-divergences, and thus no counterterms,
for $I_{7d}$ and $I_{8a}$), reflecting the aforementioned IR divergences in each case. In expressions for physical observables, the fact that IR divergences must be absent ensures that these integrals can always be eliminated in favor of the $\epsilon$-finite master integrals.
More generally, $\zeta_4$ (or equivalently $I_{6d}$) also cancels from the self-energy function contributions from massless particles in gauge theories; the amusing absence of $\zeta_n$ with even $n$ has been noted in various contexts in e.g.~\cite{Chetyrkin:1979bj,Gorishnii:1990vf,Gorishnii:1991hw,Broadhurst:1999xk,Baikov:2010je,Baikov:2010hf}.

\section{The case of all internal propagator masses equal\label{sec:allmassive}}
\setcounter{equation}{0}
\setcounter{figure}{0}
\setcounter{table}{0}
\setcounter{footnote}{1}

Next, consider the case that all of the internal propagators have the same squared mass, which will be called $t$. In this case, there are no IR divergences, so all renormalized candidate masters are $\epsilon$-finite.
By applying the IBP relations, I find that all renormalized self-energy integrals (including those with arbitrary momentum polynomials in the numerators) up to three-loop order can be written in terms of the following renormalized $\epsilon$-finite masters:
\beq
{\cal I}_1 &=& \{ A,\> B \}
,
\label{eq:allmassivebasisone}
\\ 
{\cal I}_2 &=& \{ I,\> S,\> T,\> U,\> M \} ,
\label{eq:allmassivebasistwo}
\\ 
{\cal I}_3 &=&
\{
G,\>H,\>
I_{4},\> 
I_{41},\> 
I_{411},\> 
I_{5a},\>
I_{5a1},\> 
I_{5a2},\> 
I_{5b},\>
I_{5b1},\> 
I_{5b2},\> 
I_{5c},\>
I_{6a},\> 
I_{6b},\> 
I_{6c},\>
I_{6d},\> 
I_{6d1},\> 
\nonumber \\ &&
I_{6e},\>
I_{6e1},\> 
I_{6f},\> 
I_{7a},\>
I_{7a1},\> 
I_{7a3},\>
I_{7b},\> 
I_{7b1},\>
I_{7b2},\> 
I_{7b4},\>
I_{7c},\> 
I_{7c1},\>
I_{7d},\> 
I_{7d1},\>
I_{7e},\> 
I_{7e3},\>
\nonumber \\ &&
I_{8a},\> 
I_{8b},\>
I_{8c},\> 
I_{8c1},\>
I_{8c}^{pk}
\}
.
\label{eq:allmassivebasisthree}
\eeq
The squared mass arguments are suppressed again in this section, because they are all equal. The remaining candidate master integrals 
\beq
&& \{F,\>
I_{412},\> I_{5b4},\> I_{5c1},\> I_{5c2},\> I_{5c3},\> I_{6a1},\> I_{6a2},\> I_{6a3},\> I_{6b1},\> I_{6b3},\> I_{6c1},\> I_{6c2},\> I_{6c3},\> 
\nonumber \\ &&
 I_{6c4},\> I_{6c5},\> I_{6d2},
 \> I_{6d6},\> I_{6e2},\> I_{6e3},\> I_{6e4},\> I_{6e5},\> I_{6f1},\> I_{6f5},\> I_{7a5},\> I_{7a7},\> I_{7b6},\> 
 \nonumber \\ &&
 I_{7c5},\> I_{7c6},\> I_{7d2},\> I_{7d3},\> I_{7d7},\> I_{7e1},\> I_{7e2},\> I_{7e4},\> I_{7e5},\> I_{7e6} \}
 \label{eq:allmassivesolved}
\eeq
are solved in terms of the masters, with results given in the ancillary file {\tt Iallmassive}. The derivatives $\displaystyle s \frac{d}{ds}$ of the masters are also provided in an ancillary file {\tt Iallmassivesdds}. The choice of masters above is somewhat arbitrary, but has been made in such a way as to make denominators simple, with factors of $s-4t$, $s-9t$, and $s-16 t$
corresponding to the threshold singularities from 2-particle, 3-particle, and 4-particle cuts respectively. However, also present in a few cases in the expressions for $\displaystyle s \frac{d}{ds}$ of the masters and for the solved integrals 
are denominator factors $s-t$, $s-3t$, $s + 8t$, and
$s^2 - 8 s t + 4 t^2$, which do not correspond to true thresholds. (These denominator factors could be eliminated at the expense of increasing the set of masters to a larger overcomplete set with some algebraic identities relating them, but there is no great advantage gained by doing so.)

It is now straightforward to obtain series solutions to the first-order differential equation in $s$, using boundary conditions given at $s=0$ by the known vacuum integrals,
\beq
I(t,t,t) &=& t \left (3 \cI - \frac{15}{2} + 6 L_t - \frac{3}{2} L_t^2 \right )
,
\\
F(t,t,t,t) &=& t \left ( \frac{53}{12} + \frac{13}{4} L_t - 4 L_t^2 + L_t^3 \right )
,
\\
G(t,t,t,t,t) &=& t \left [12 \cI -\frac{97}{3} + 6 \zeta_3 +
\left (26 - 6 \cI \right ) L_t - 8 L_t^2 + L_t^3 \right ] 
,
\\
H(t,t,t,t,t,t) &=& \conHsix + 6 \zeta_3 (1 - L_t),
\eeq
with
\beq
\cI &\equiv& \sqrt{3}\, {\rm Im}\left [{\rm Li}_2(e^{2\pi i/3}) \right ]
\>=\> 1.1719536193\ldots
,
\\
\conHsix &\equiv& 16\, {\rm Li}_4(1/2) - 17 \zeta_4 + \frac{2}{3} \ln^2 (2) [\ln^2(2) - \pi^2] - 3 \cI^2
\>=\> 17.2476198987\ldots
.
\eeq
Defining $r=s/t$, I have obtained power series solutions convergent
for $|r| < 4$. (The physical reason for this range of convergence is that the point $r=4$ corresponds to the lowest 2-particle cut threshold.) The series results up to order $r^{36}$,
for the masters in eqs.~(\ref{eq:allmassivebasisone})-(\ref{eq:allmassivebasisthree}) as well as the solved integrals in eq.~(\ref{eq:allmassivesolved}), are given in the ancillary file {\tt Iallmassiveseries}. The coefficients in these series involve only rational numbers and the constants $\zeta_3$, $\cI$, and $\conHsix$. The only appearances of the constant $\conHsix$ are in $H$ itself and in the $r^0$ term in the expansion of $I_{6d}$.

\begin{figure}[!t]
  \begin{minipage}[]{0.495\linewidth}
    \includegraphics[width=7.cm,angle=0]{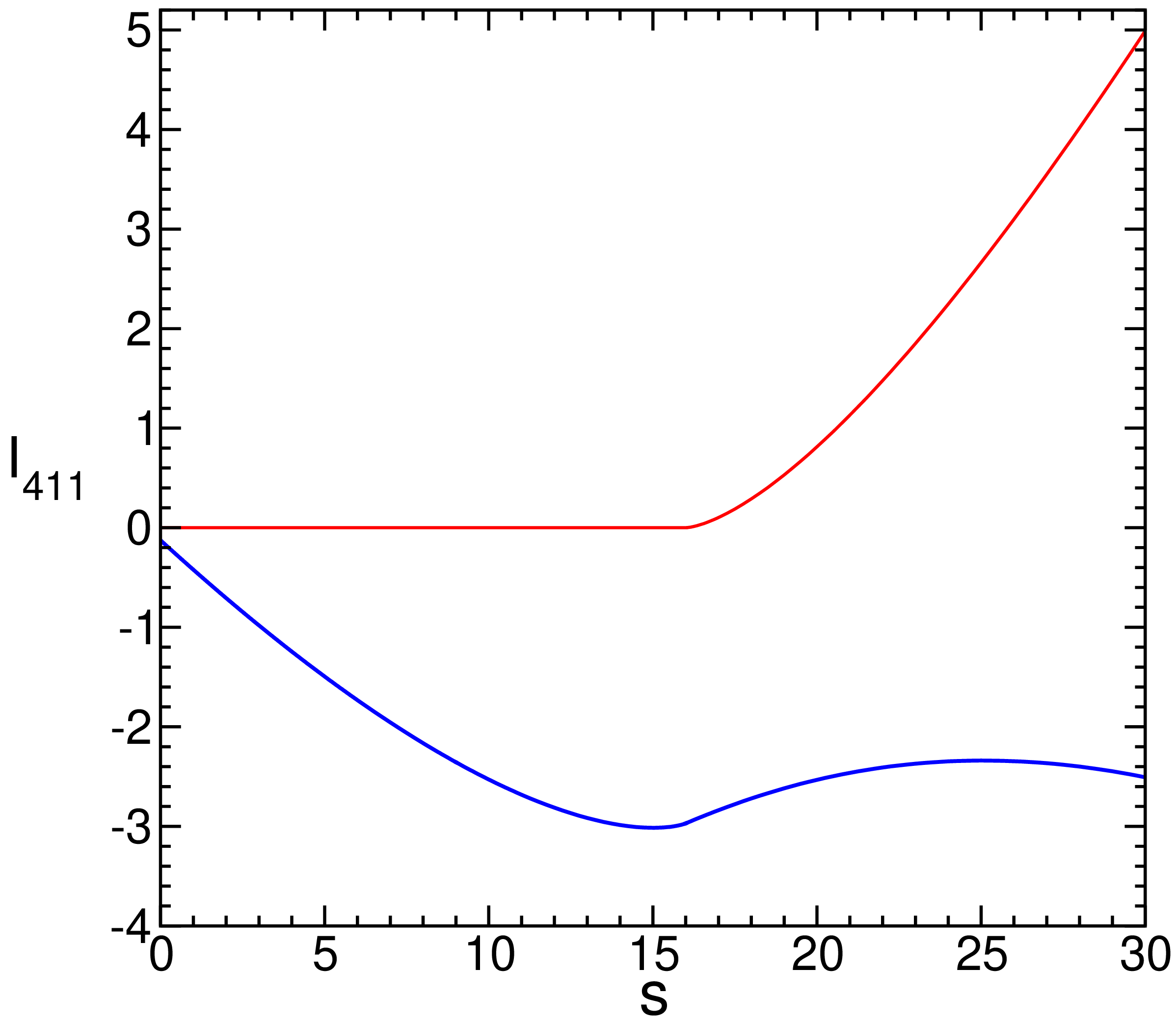}
  \end{minipage}
  \begin{minipage}[]{0.495\linewidth}
    \includegraphics[width=7.cm,angle=0]{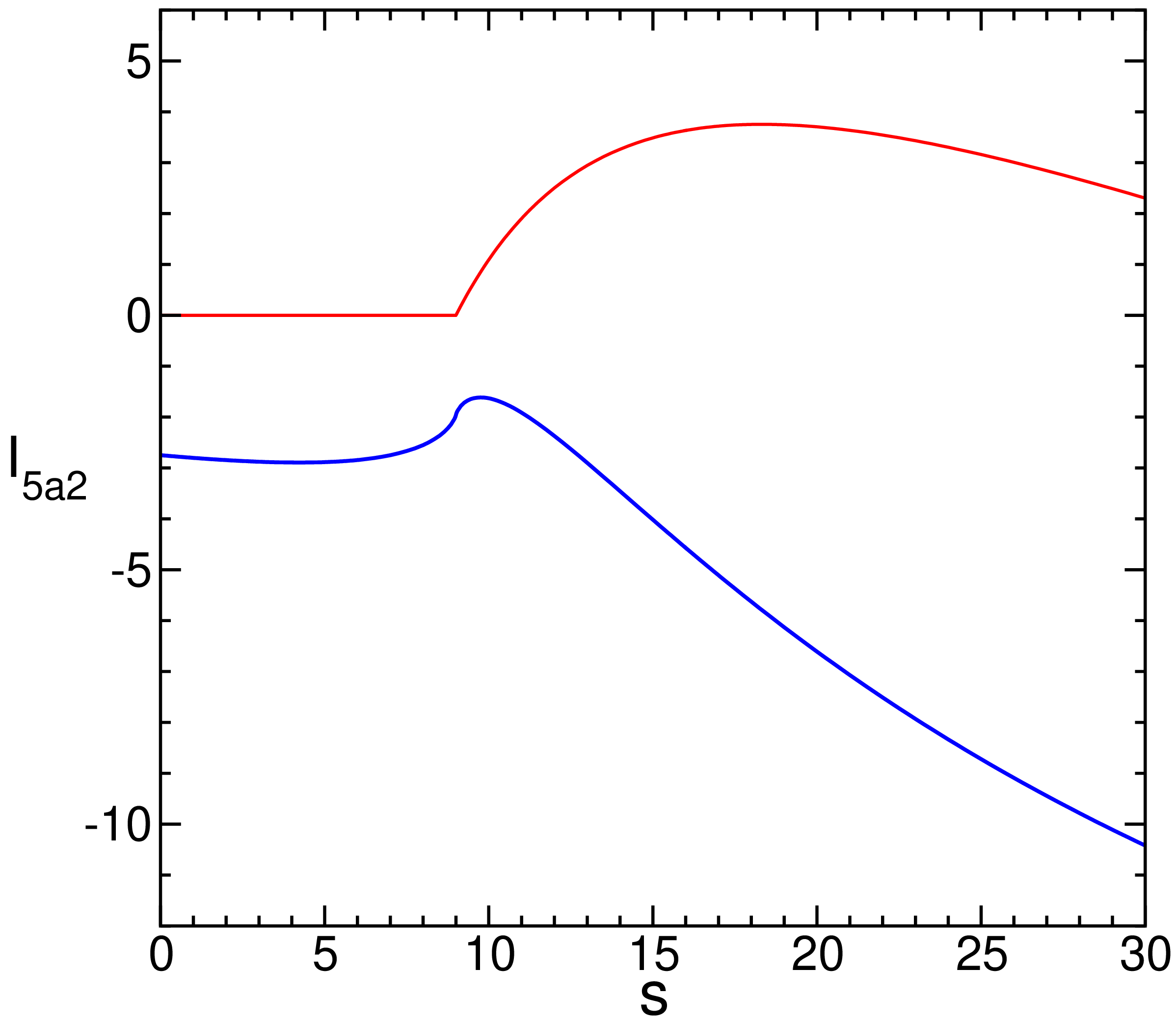}
  \end{minipage}
  \begin{minipage}[]{0.495\linewidth}
    \includegraphics[width=7.cm,angle=0]{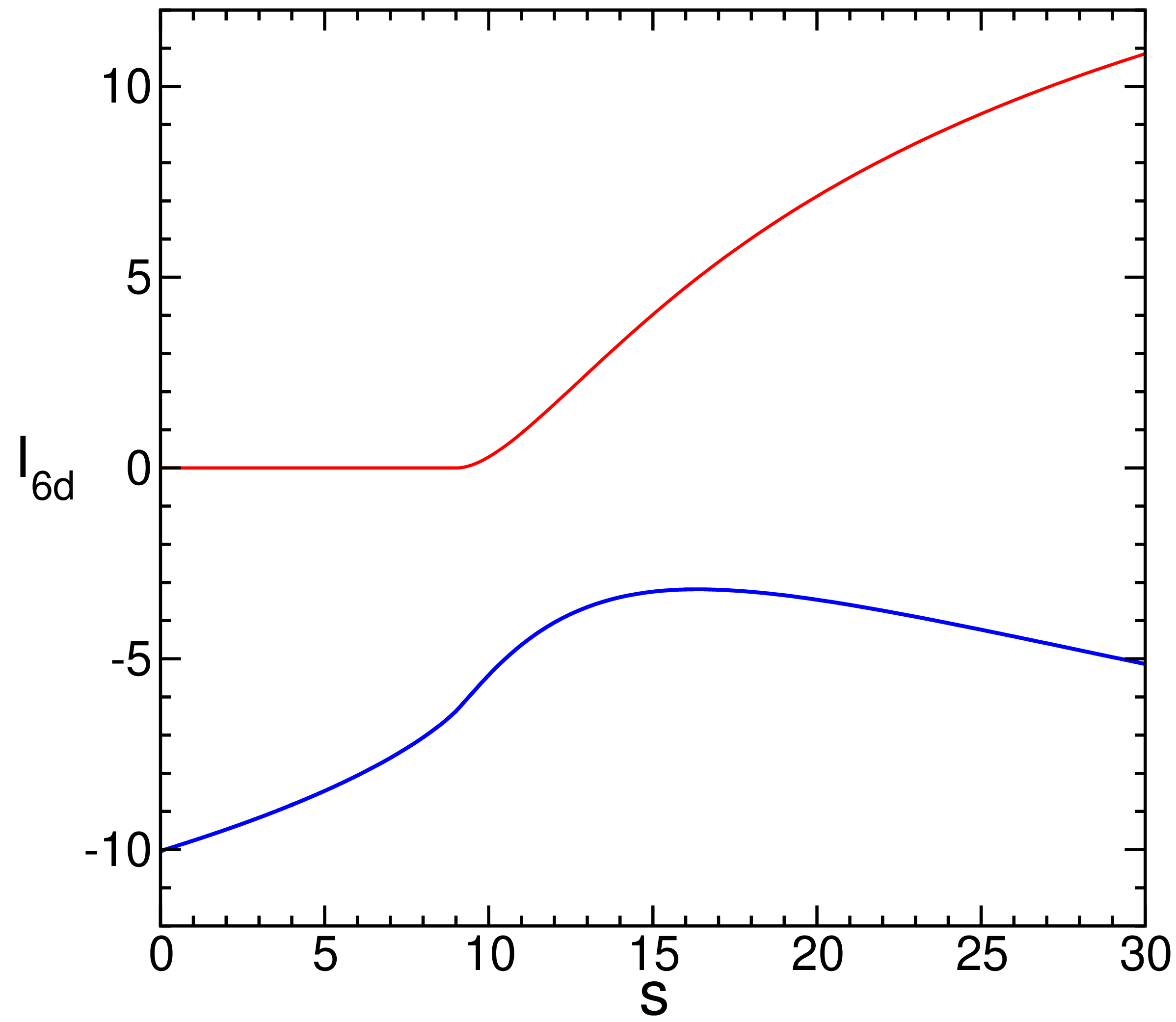}
  \end{minipage}
  \begin{minipage}[]{0.495\linewidth}
    \includegraphics[width=7.cm,angle=0]{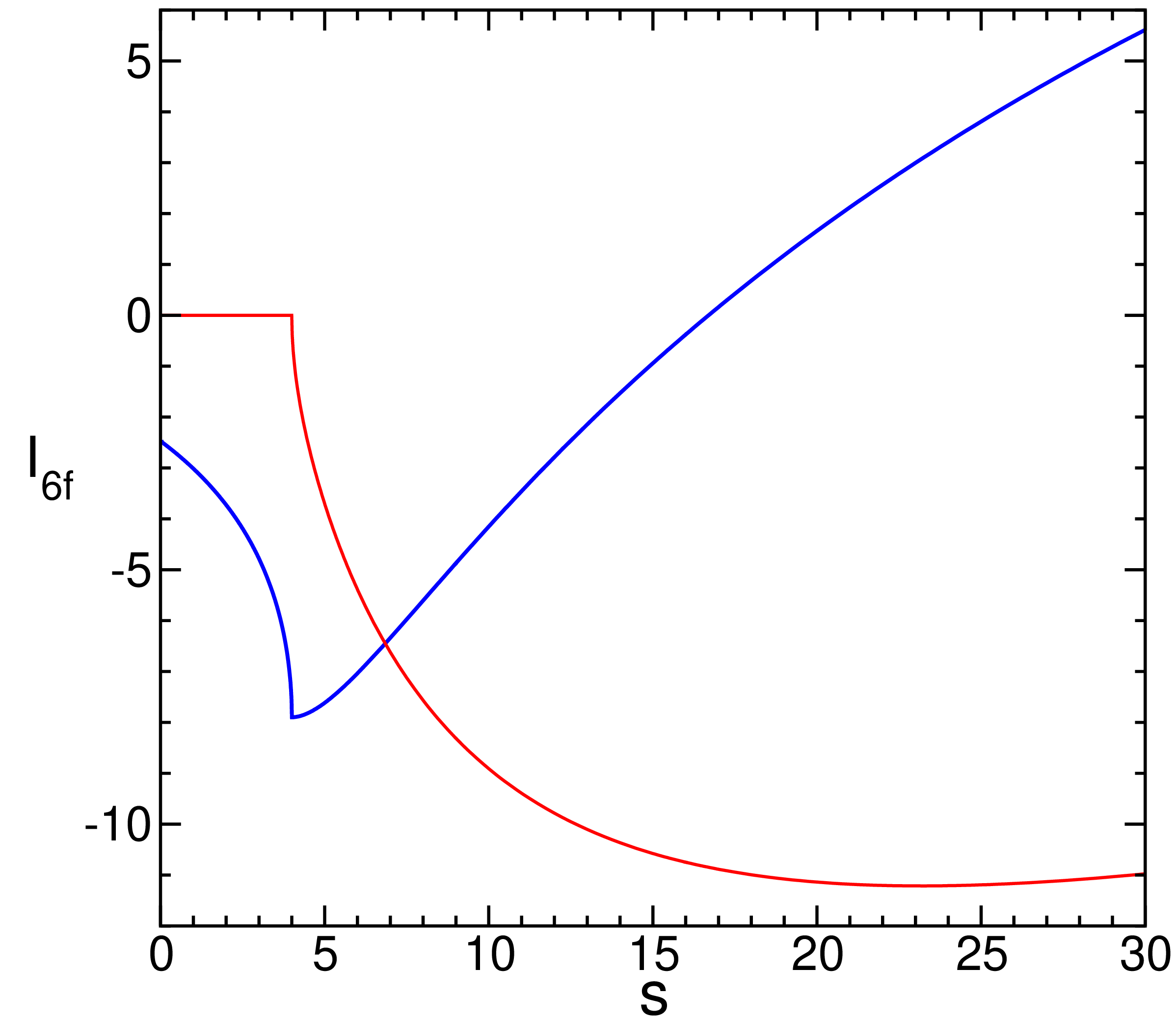}
  \end{minipage}
  \begin{center}
    \begin{minipage}[]{0.95\linewidth}
\caption{\label{fig:integraldataallmassive}Sample results, for the dimensionless integrals $I_{411}$, $I_{5a2}$, $I_{6d}$, and $I_{6f}$, with all propagator squared masses and the renormalization scale $Q$ set equal to unity ($t= Q^2 = 1$), as a function of the external momentum invariant $s$. The results were obtained by numerical solution of the coupled first-order differential equations in $s$ as provided in the ancillary file {\tt Iallmassivesdds}, starting from the series solution provided in the file {\tt Iallmassiveseries}. In each case, the blue (heavier) line is the real part, and the red (lighter) line is the imaginary part. The lowest threshold is at $s=16$ (four-particle cut) for $I_{411}$, at $s=9$ (three-particle cut) for $I_{5a2}$ and $I_{6d}$, and at $s=4$ (two-particle cut) for $I_{6f}$.}
    \end{minipage}
  \end{center}
\end{figure}

For general $s$ not necessarily small compared to $4t$, a numerical integration of the coupled first order differential equations for the integrals in eqs.~(\ref{eq:allmassivebasisone})-(\ref{eq:allmassivebasisthree}), starting from the series solution at $s=0.5$ (for example) as the initial condition, is sufficient to quickly obtain accurate numerical results, as explained in section \ref{subsec:numericaleval}. As a check, I have verified numerically that the results for $s\gg t$ indeed asymptotically approach those given in eqs.~(\ref{eq:Ballmassless})-(\ref{eq:I8cpkallmassless})
in the previous section. 
A few examples of results for dimensionless (6-propagator) integrals as a function of $s$ are shown in Figure \ref{fig:integraldataallmassive}, for the case $t=Q^2 = 1$.
The integral $I_{6f}$ has a two-particle cut threshold at $s=4$, leading to cuspy behavior near that point. The integrals $I_{5a2}$ and $I_{6d}$ have three-particle cut thresholds at $s=9$. The integral $I_{411}$ has a relatively smooth four-particle cut threshold at $s=16$. (The large $s$ asymptotic limits of the previous section for $I_{6d}$ and $I_{6f}$ are accurately realized only for $s$ much larger than the ranges shown in the figures.)

Another way of obtaining numerical results for general $s$ for integrals with a single internal mass scale is to make series expansions (in general, with square root and logarithmic factors) about the threshold points $s = 4t$ (two massive particle cut) and $9t$ (three massive particle cut) and $16t$ (four massive particle cut) and $\infty$.  The coefficients in these series expansions can then be determined by matching at points within the common range of convergence of pairs of series, starting from the analytically known series coefficients for the expansion about $s=0$. However, it does  not seem so easy to generalize this method to the case of arbitrary different internal propagator squared masses for all $s$, and so the results will not be pursued here. 

\FloatBarrier
 
\section{Integrals with odd thresholds \label{sec:odd}}
\setcounter{equation}{0}
\setcounter{figure}{0}
\setcounter{table}{0}
\setcounter{footnote}{1}

In any unbroken gauge theory (such as QED or QCD) with massive and massless fermions and massless gauge bosons, the allowed interaction vertices have an even number of massive lines. Consider a self-energy diagram topology with a single internal propagator squared mass scale called $t$ (in honor of the top quark), with the other internal propagators massless. The cuts of the diagram will correspond to thresholds at $s = n^2 t$, where the $n$ are either all even integers $n=0,2,4$, or else all odd integers $n=1,3$. Furthermore, it is easy to see that all descendants of the diagram obtained by removing internal propagators will have the same property. 

\begin{figure}[!t]
\begin{center}
\includegraphics[width=11.25cm,angle=0]{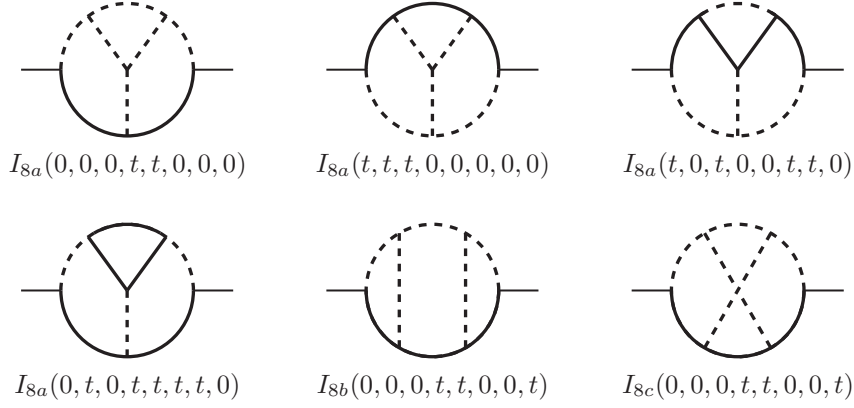}
\end{center}
\vspace{-0.3cm}
\begin{minipage}[]{0.95\linewidth}
\caption{\label{fig:oddthresholdtopologies}
The odd-threshold single-mass three-loop self-energy topologies considered (along with their descendants) in section \ref{sec:odd}. The heavy solid internal lines represent propagators with squared mass $t$, and the dashed lines represent massless propagators.}
\end{minipage}
\end{figure}

In this section, I consider three-loop self energy integrals  with possible thresholds only at $s=t$ and/or $s=9 t$, which are referred to here as ``odd threshold" integrals. These are the ones that can arise in QCD corrections to the self energies of the $W$ boson in which the $W$ boson couples to $t,b$, with the bottom quark treated as massless. They arise from diagram topologies corresponding to the scalar integrals 
\beq
&&
I_{8a}(0,0,0,t,t,0,0,0),\>\>\>
I_{8a}(t,t,t,0,0,0,0,0),\>\>\>
I_{8a}(t,0,t,0,0,t,t,0),\>\>\>
\nonumber \\ &&
I_{8a}(0,t,0,t,t,t,t,0),\>\>\>
I_{8b}(0,0,0,t,t,0,0,t),\>\>\>
I_{8c}(0,0,0,t,t,0,0,t),\>\>\>
\label{eq:oddthresholdtopologies}
\eeq
as depicted in Figure \ref{fig:oddthresholdtopologies},
and their descendants obtained by removing internal lines in all possible ways.
Note that these six 8-propagator topologies are linked by a variety of common descendants.

Applying the IBP identities, I find that all scalar self-energy functions with these topologies can be expressed in terms of the one-loop, two-loop, and three-loop renormalized $\epsilon$-finite master integrals:
\beq
{\cal I}^{(1)} &=& \{\, A(t),\> B(0, t)\,\} ,
\label{eq:oddthreshbasisone}
\\
{\cal I}^{(2)} &=& 
\{\, S(0, 0, t),\> S(t, t, t),\> U(t, 0, t, t),\>M(0, 0, t, t, 0) \,\} ,
\\
{\cal I}^{(3)} &=&
\{
H(0, 0, t, 0, t, t),\> 
H(0, t, t, t, 0, t),\>
I_{4}(0, t, t, t),\> 
I_{5a}(t, 0, 0, t, t),\> 
I_{5b1}(t, t, t, 0, t),\>
\nonumber \\ &&
I_{6a}(t, 0, 0, 0, t, t),\> 
I_{6b1}(t, 0, 0, t, t, t),\> 
I_{6d}(0, 0, t, 0, t, 0),\>
I_{6d}(0, 0, t, t, 0, t),\> 
\nonumber \\ &&
I_{6d}(t, 0, 0, 0, 0, 0),\> 
I_{6d}(t, 0, 0, t, t, t),\>
I_{6d}(t, t, t, t, t, 0),\> 
I_{6d1}(t, t, t, t, t, 0),\> 
\nonumber \\ &&
I_{6e}(0, 0, t, t, t, t),\>
I_{6e}(0, t, 0, 0, 0, t),\> 
I_{6e}(t, 0, t, 0, 0, t),\> 
I_{6e}(t, t, 0, t, t, t),\>
I_{6f}(0, t, t, 0, 0, t),\> 
\nonumber \\ &&
I_{6f1}(t, 0, 0, t, 0, t),\> 
I_{7a}(0, 0, 0, 0, t, t, t),\>
I_{7a}(0, 0, t, t, 0, 0, 0),\> 
I_{7a}(0, t, t, 0, 0, t, 0),\>
\nonumber \\ &&
I_{7a}(t, t, 0, 0, t, t, 0),\> 
I_{7a}(t, t, t, t, 0, 0, t),\>
I_{7a5}(t, t, 0, 0, t, t, 0),\> 
I_{7b}(0, 0, t, 0, t, 0, 0),\>
\nonumber \\ &&
I_{7b}(t, 0, 0, 0, t, 0, t),\> 
I_{7b}(t, t, t, 0, t, t, 0),\>
I_{7c}(0, 0, t, t, 0, 0, 0),\> 
I_{7d}(0, t, 0, t, 0, t, 0),\>
\nonumber \\ &&
I_{7d}(0, t, 0, t, t, 0, t),\> 
I_{7e}(0, t, t, 0, 0, 0, 0),\>
I_{7e}(0, t, t, 0, 0, t, t),\> 
I_{8a}(0, t, 0, t, t, t, t, 0),\>
\nonumber \\ &&
I_{8b}(0, 0, 0, t, t, 0, 0, t),\> 
I_{8c}(0, 0, 0, t, t, 0, 0, t),\>
I_{8c}^{pk}(t, t, t, 0, 0, 0, 0, 0)
\}
.
\label{eq:oddthreshbasisthree}
\eeq
For the other candidate master integrals, the solutions (obtained from repeated use of the IBP relations) in terms of the masters above are given in an ancillary file {\tt Iodd}. The complete\footnote{Although they do not have doubled massless propagators, $I_{8a}(0,0,0,t,t,0,0,0)$ and $I_{7d}(t,0,0,0,0,0,0)$ are IR-divergent, and are not candidates for renormalized $\epsilon$-finite master integrals.}
 list of
such solved renormalized $\epsilon$-finite candidate master integrals is:
\beq
&& 
\{
I(0, t, t),\>
T(t, 0, 0),\> 
T(t, t, t),\> 
U(0, t, 0, t),\> 
U(t, 0, 0, 0),\>
F(t, 0, 0, t),\> 
F(t, t, t, t),\> 
\nonumber \\ &&
G(0, 0, 0, t, t),\> 
G(0, t, t, t, t),\> 
G(t, 0, t, 0, t),\>  
I_{4}(0, 0, 0, t),\> 
I_{41}(t, 0, 0, 0),\> 
I_{41}(t, 0, t, t),\> 
\nonumber \\ &&
I_{411}(t, 0, 0, 0),\> 
I_{411}(t, 0, t, t),\> 
I_{412}(t, t, 0, t),\> 
I_{5a}(0, 0, t, 0, t),\> 
I_{5a}(t, 0, 0, 0, 0),\> 
I_{5a}(t, t, t, t, t),\> 
\nonumber \\ &&
 I_{5a1}(t, 0, 0, 0, 0),\> 
 I_{5a1}(t, 0, 0, t, t),\> 
 I_{5a1}(t, t, t, t, t),\> 
 I_{5a2}(0, t, 0, 0, t),\> 
 I_{5a2}(t, t, t, 0, 0),\> 
\nonumber \\ &&
 I_{5a2}(t, t, t, t, t),\> 
 I_{5b}(0, 0, t, 0, 0),\> 
 I_{5b}(0, 0, t, t, t),\> 
 I_{5b}(t, 0, 0, 0, t),\> 
 I_{5b}(t, t, t, 0, t),\> 
\nonumber \\ &&
 I_{5b1}(t, 0, 0, 0, t),\> 
 I_{5b2}(0, t, 0, 0, 0),\> 
 I_{5b2}(0, t, 0, t, t),\> 
 I_{5b2}(t, t, t, 0, t),\> 
 I_{5b4}(0, 0, t, t, t),\> 
\nonumber \\ &&
 I_{5b4}(t, 0, 0, t, 0),\> 
 I_{5b4}(t, t, t, t, 0),\> 
 I_{5c}(0, t, 0, 0, t),\> 
 I_{5c}(0, t, t, t, t),\> 
 I_{5c}(t, 0, 0, 0, 0),\> 
 \nonumber \\ &&
 I_{5c}(t, 0, 0, t, t),\> 
 I_{5c3}(0, t, t, 0, 0),\> 
 I_{5c3}(0, t, t, t, t),\> 
 I_{5c3}(t, 0, t, 0, t),\> 
 I_{5c2}(0, t, 0, 0, t),\> 
\nonumber \\ &&
 I_{5c2}(0, t, t, t, t),\> 
 I_{5c1}(t, 0, 0, 0, 0),\> 
 I_{5c1}(t, 0, 0, t, t),\> 
 I_{6a}(0, t, 0, t, 0, t),\> 
 I_{6a}(t, 0, 0, 0, 0, 0),\> 
\nonumber \\ && 
 I_{6a1}(t, 0, 0, 0, 0, 0),\> I_{6a1}(t, 0, 0, 0, t, t),\> I_{6a2}(0, t, 0, t, 0, t),\> 
 I_{6a3}(0, t, t, 0, 0, t),\> I_{6a3}(t, 0, t, t, 0, 0), \> 
\nonumber \\ &&
I_{6b}(0, t, 0, 0, 0, t),\> 
 I_{6b}(0, t, t, t, 0, t),\> I_{6b1}(t, 0, 0, t, 0, 0),\> I_{6b3}(0, t, t, t, 0, t),\> 
 I_{6b3}(t, 0, t, 0, 0, 0),\> 
\nonumber \\ && 
 I_{6b3}(t, 0, t, 0, t, t),\> 
 I_{6c}(0, t, 0, t, 0, t),\> 
 I_{6c}(0, t, t, 0, 0, 0),\> 
 I_{6c}(0, t, t, 0, t, t),\> 
 I_{6c}(t, 0, 0, 0, 0, 0),\> 
\nonumber \\ && 
 I_{6c}(t, 0, 0, 0, t, t),\> 
 I_{6c}(t, 0, t, t, 0, t),\> 
 I_{6c1}(t, 0, 0, 0, 0, 0),\> 
 I_{6c1}(t, 0, 0, 0, t, t),\> 
 I_{6c1}(t, 0, t, t, 0, t),\> 
\nonumber \\ &&
 I_{6c2}(0, t, 0, t, 0, t),\> 
 I_{6c2}(0, t, t, 0, 0, 0),\> 
 I_{6c2}(0, t, t, 0, t, t),\> 
 I_{6c3}(0, t, t, 0, 0, 0),\> 
 I_{6c3}(0, t, t, 0, t, t),\> 
\nonumber \\ &&
 I_{6c3}(t, 0, t, t, 0, t),\> I_{6c4}(0, t, 0, t, 0, t),\> 
 I_{6c4}(t, 0, t, t, 0, t),\> I_{6c5}(0, t, 0, t, t, 0),\> I_{6c5}(0, t, t, 0, t, t),\> 
\nonumber \\ &&
 I_{6c5}(t, 0, 0, 0, t, t),\> I_{6c5}(t, 0, t, t, t, 0),\> I_{6d1}(t, 0, 0, 0, 0, 0),\> 
 I_{6d1}(t, 0, 0, t, t, t),\> I_{6d2}(0, t, 0, 0, t, t),\> 
\nonumber \\ &&
 I_{6d2}(0, t, 0, t, 0, 0),\> 
 I_{6d2}(t, t, t, 0, 0, t),\> I_{6d2}(t, t, t, t, t, 0),\> I_{6d6}(0, 0, t, t, 0, t),\> 
 I_{6d6}(t, 0, 0, t, t, t),\> 
\nonumber \\ &&
 I_{6e}(0, 0, t, t, 0, 0),\> I_{6e1}(t, 0, t, 0, 0, t),\> 
 I_{6e1}(t, t, 0, t, t, t),\> I_{6e5}(0, 0, t, t, t, t),\> I_{6e5}(0, t, 0, 0, t, 0),\> 
\nonumber \\ &&
 I_{6e5}(t, 0, t, 0, t, 0),\> I_{6e5}(t, t, 0, t, t, t),\> I_{6e2}(0, t, 0, 0, 0, t),\> 
 I_{6e2}(t, t, 0, t, t, t),\> I_{6e4}(0, 0, t, t, 0, 0),\> 
\nonumber \\ &&
 I_{6e4}(0, 0, t, t, t, t),\> 
 I_{6e4}(t, t, 0, t, t, t),\> I_{6e3}(0, 0, t, t, 0, 0),\> I_{6e3}(0, 0, t, t, t, t),\> 
 I_{6e3}(t, 0, t, 0, 0, t),\> 
\nonumber \\ &&
 I_{6f}(0, t, 0, t, 0, 0),\> I_{6f}(0, t, 0, t, t, t),\> 
 I_{6f5}(0, t, 0, t, t, t),\> I_{6f5}(0, t, t, 0, t, 0),\> I_{6f1}(t, 0, t, 0, 0, 0),\> 
\nonumber \\ &&
 I_{6f1}(t, 0, t, 0, t, t),\> I_{7a1}(t, 0, 0, t, t, 0, 0),\> 
 I_{7a1}(t, t, 0, 0, t, t, 0),\> I_{7a1}(t, t, t, t, 0, 0, t),\> 
\nonumber \\ &&
 I_{7a3}(0, 0, t, t, 0, 0, 0),\> I_{7a3}(0, t, t, 0, 0, t, 0),\> 
 I_{7a3}(t, t, t, t, 0, 0, t),\> I_{7a5}(0, 0, 0, 0, t, t, t),\> 
\nonumber \\ &&
 I_{7a5}(t, 0, 0, t, t, 0, 0),\> I_{7a7}(0, 0, 0, 0, t, t, t),\> 
 I_{7a7}(t, t, t, t, 0, 0, t),\> I_{7b1}(t, 0, 0, 0, t, 0, t),\> 
\nonumber \\ &&
 I_{7b1}(t, t, t, 0, t, t, 0),\> I_{7b2}(0, t, 0, t, 0, 0, 0),\> 
 I_{7b2}(t, t, t, 0, t, t, 0),\> I_{7b2}(t, t, t, t, 0, 0, t),\> 
\nonumber \\ &&
 I_{7b4}(0, t, 0, t, 0, 0, 0),\> I_{7b4}(t, 0, 0, t, 0, t, 0),\> 
 I_{7b4}(t, t, t, t, 0, 0, t),\> I_{7b6}(t, 0, 0, t, 0, t, 0),\> 
\nonumber \\ &&
 I_{7b6}(t, t, t, 0, t, t, 0),\> I_{7c}(0, 0, t, t, 0, t, t),\> 
 I_{7c1}(t, t, 0, 0, 0, 0, 0),\> I_{7c1}(t, t, 0, 0, 0, t, t),\> 
\nonumber \\ &&
 I_{7c6}(0, 0, t, t, 0, t, t),\> I_{7d}(t, 0, 0, 0, t, t, t),\> 
 I_{7d1}(t, 0, 0, 0, t, t, t),\> I_{7d2}(0, t, 0, t, 0, t, 0),\> 
\nonumber \\ &&
 I_{7d2}(0, t, 0, t, t, 0, t),\> I_{7d3}(0, t, t, 0, 0, t, t),\> 
 I_{7d3}(0, t, t, 0, t, 0, 0),\> I_{7d3}(t, 0, t, t, 0, 0, t),\> 
\nonumber \\ &&
 I_{7d7}(0, t, 0, t, t, 0, t),\> I_{7d7}(t, 0, 0, 0, t, t, t),\> 
 I_{7e}(t, 0, 0, t, 0, 0, t),\> I_{7e1}(t, 0, 0, t, 0, 0, t),\> 
\nonumber \\ &&
 I_{7e2}(0, t, t, 0, 0, 0, 0),\> I_{7e2}(0, t, t, 0, 0, t, t),\> 
 I_{7e3}(0, t, t, 0, 0, 0, 0),\> I_{7e3}(0, t, t, 0, 0, t, t),\> 
\nonumber \\ &&
 I_{7e4}(t, 0, 0, t, 0, 0, t),\> I_{7e6}(0, t, t, 0, 0, t, t),\> 
 I_{7e6}(t, 0, 0, t, 0, t, 0),\> I_{8a}(t, 0, t, 0, 0, t, t, 0),\> 
\nonumber \\ &&
 I_{8a}(t, t, t, 0, 0, 0, 0, 0) \> \} 
 .
 \label{eq:oddbasissolved}
\eeq
Furthermore, the derivatives of the master integrals in eqs.~(\ref{eq:oddthreshbasisone})-(\ref{eq:oddthreshbasisthree}) with respect to $s$ can be expressed in terms of that same set of masters.
The results for the derivatives $s \frac{d}{ds}$ acting on all of the master integrals  are provided in the ancillary file {\tt Ioddsdds}.

\baselineskip=15.2pt

For practical numerical evaluation, I have derived the series solutions to the differential equations in $r =s/t$, convergent for $|r|<1$, using as boundary conditions the limits of the integrals at $s=0$. These boundary conditions 
involve known vacuum integrals, using the notation of ref.~\cite{Martin:2016bgz}:
\vspace{-0.2cm}
\beq
\overline{F}(0,t,t,t) &=& t \left [ \conFbar + (3 \cI - 3/2) L_t + \frac{3}{2} L_t^2 - \frac{1}{2} L_t^3 \right ],
\\
G(t,0,0,t,t) &=& t \left [\conG + 
\left (15 + \zeta_2 - 3 \cI \right ) L_t
- 5 L_t^2 + \frac{2}{3} L_t^3 
\right ]
,
\\
H(0,t,t,t,0,t) &=& \conHfour + 6 \zeta_3 (1 - L_t)
,
\label{eq:H0ttt0t}
\\
H(0,t,t,t,t,t) &=& \conHfive + 6 \zeta_3 (1 - L_t)
,
\eeq
\vspace{-0.8cm}

\noindent which involve the numerical constants
\vspace{-0.3cm}
\beq
\conFbar &=& \frac{1}{2} 
-3 \cI
+6 \sqrt{3} (\ln 3 - \Lsthree) - \frac{\pi^3}{\sqrt{3}}
\>\approx\>  9.0968675373
,
\\
\conG &=& -\frac{52}{3} + 6 \cI - \frac{\pi^2}{3} -
\frac{2 \pi^3}{9\sqrt{3}}
- \frac{4}{3}\zeta_3
\>\approx\> -19.1723294414
,
\\
\conHfour &=& 
32\, {\rm Li}_4(1/2) 
-\frac{11 \pi^4}{45} 
+ \frac{4}{3} \ln^2(2) \left [ \ln^2(2) - \pi^2 \right ]
\>\approx\>
-13.2665092775
,
\\
\conHfive &=& \frac{7 \pi^4}{30} - 2 \cI^2 
+ 4 \pi \Lsthree
- 6 \Lspfour
- \frac{26}{3} \ln(3) \zeta_3
\>\approx\> -15.4292012365
,
\eeq
where
\vspace{-0.45cm}
\beq
\Lsthree &\equiv&  
-\int_0^{2\pi/3} dx \ln^2 [2 \sin(x/2)] \>\approx\>
-2.1447672125694944 ,
\\
\Lspfour &\equiv&  
-\int_0^{2\pi/3} dx \,x \ln^2 [2 \sin(x/2)] \>\approx\>
-0.4976755516066472 .
\eeq
\vspace{-0.6cm}

\noindent 
Note that $\overline{F}(0,t,t,t)$ is defined in ref.~\cite{Martin:2016bgz} as
\vspace{-0.3cm}
\beq
\overline{F}(0,t,t,t) &=& 
\lim_{x \rightarrow 0} \left[ F(x,t,t,t) + L_x I(t,t,t) \right ]
,
\eeq 
and can also be evaluated in terms of other renormalized $\epsilon$-finite vacuum integrals as
\beq
\overline{F}(0,t,t,t) &=&
\frac{1}{3} (1 + L_t) I(t,t,t)
+ t \left [\frac{19}{3} - L_t - 2 L_t^2 + \frac{2}{3} L_t^3
-2 \frac{d}{dx}G(x,t,t,0,t) \Bigl |_{x=t}
\right ].
\phantom{xxx}
\eeq

The series expansions for the renormalized $\epsilon$-finite three-loop odd-threshold self-energy integrals are given in the ancillary file {\tt Ioddseries} in terms of powers of $r=s/t$
up to $r^{36}$, with coefficients that
involve $L_t$, $c_F$, $c_G$, $c_H'$, $c_H''$ as well as $\zeta_2$, $\zeta_3$, $\zeta_4$ and rational numbers. These series converge for $|r|<1$, and easily suffice for the relevant Standard Model physical value $r = m_W^2/m_t^2$.

\begin{figure}[!t]
  \begin{minipage}[]{0.495\linewidth}
    \includegraphics[width=7.cm,angle=0]{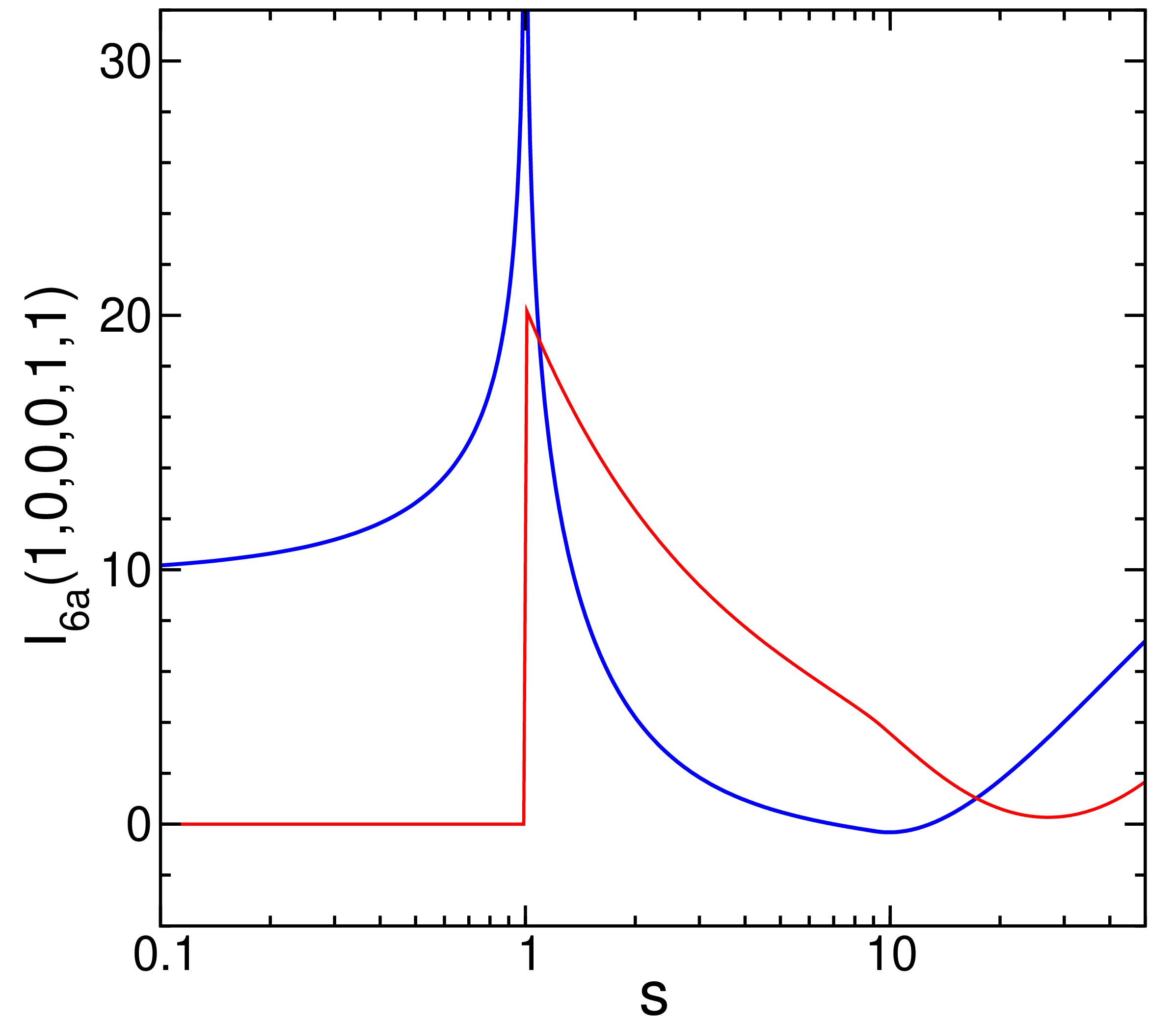}
  \end{minipage}
  \begin{minipage}[]{0.495\linewidth}
    \includegraphics[width=7.cm,angle=0]{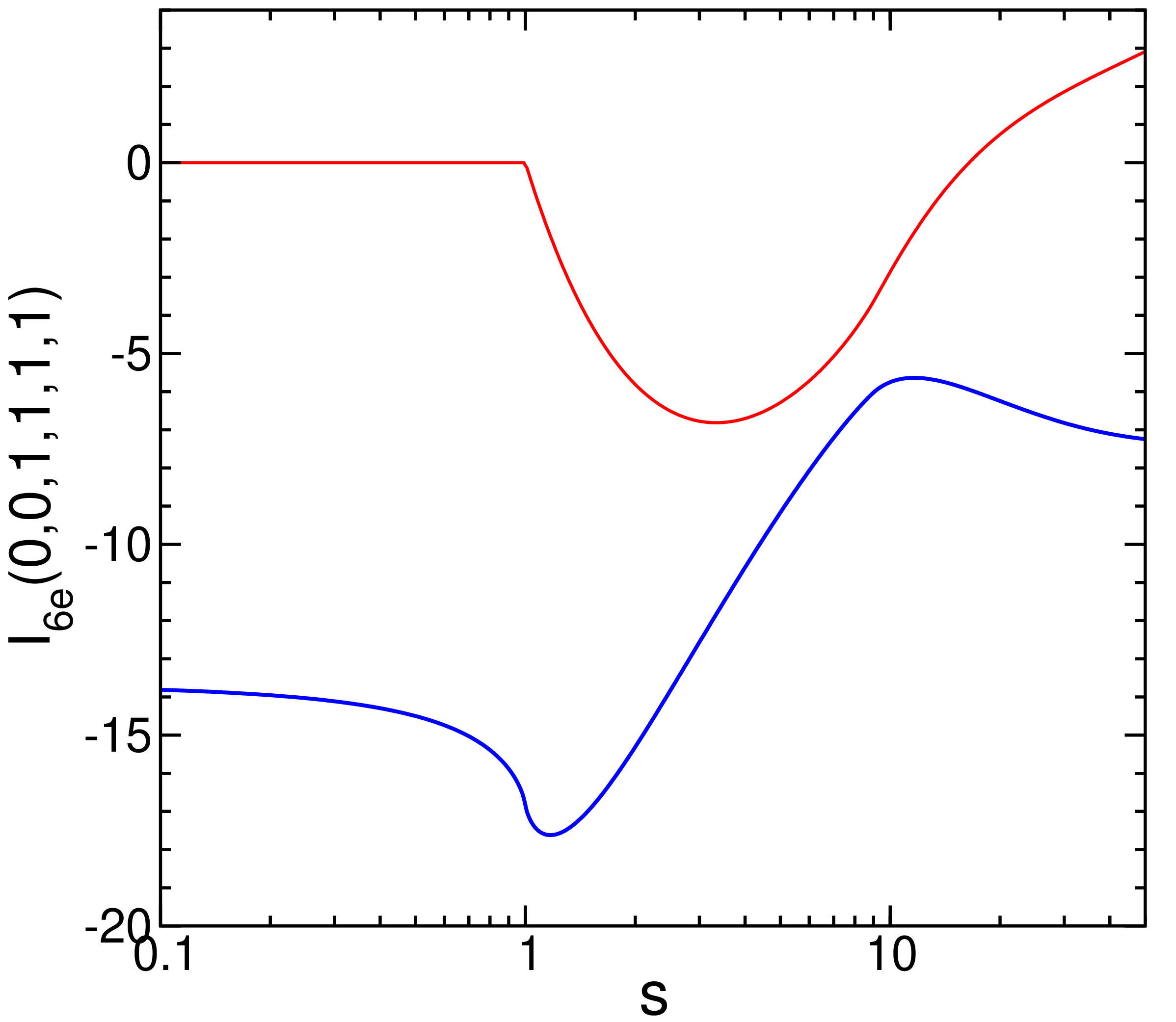}
  \end{minipage}
  \begin{minipage}[]{0.495\linewidth}
    \includegraphics[width=7.cm,angle=0]{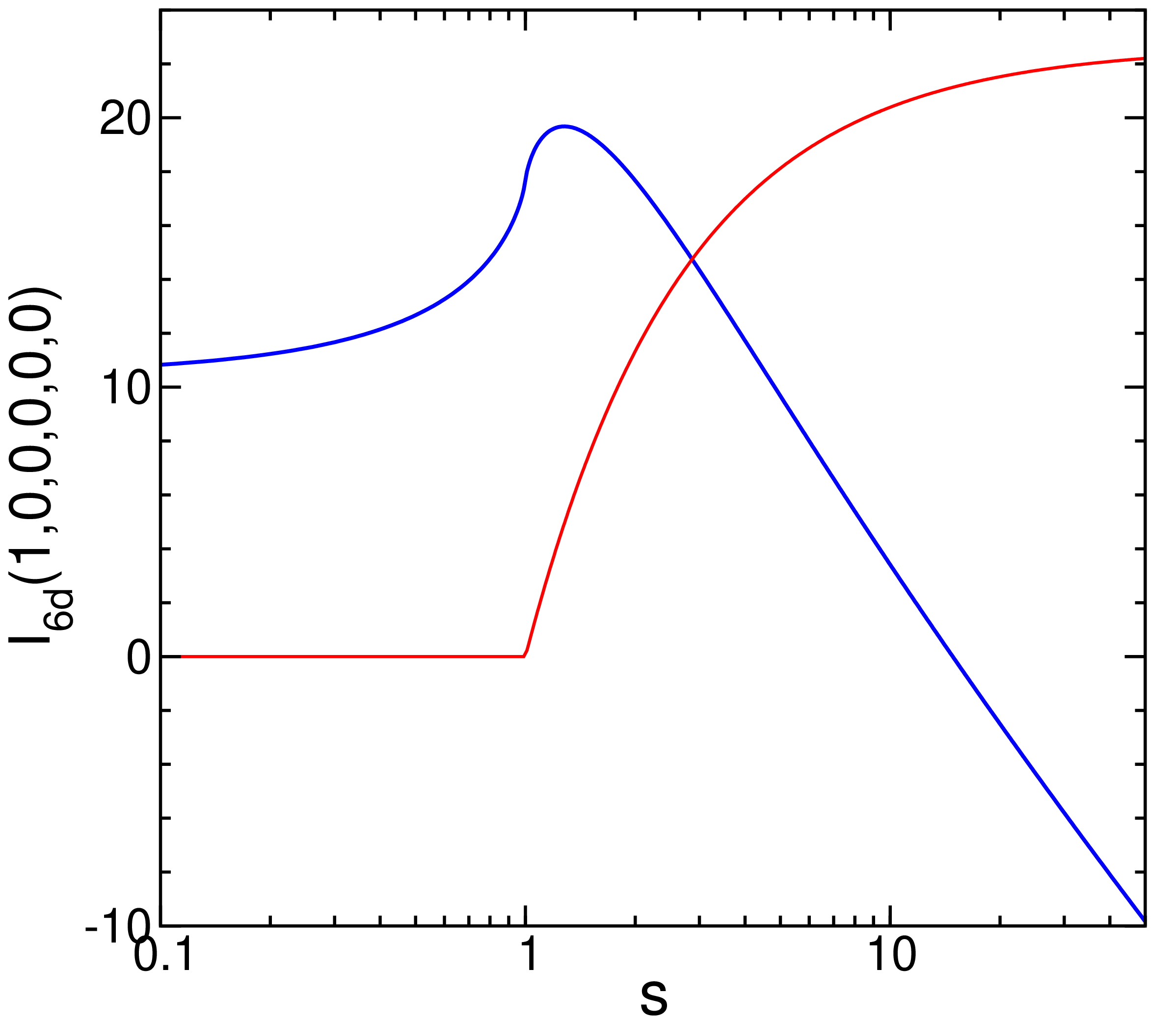}
  \end{minipage}
  \begin{minipage}[]{0.495\linewidth}
    \includegraphics[width=7.cm,angle=0]{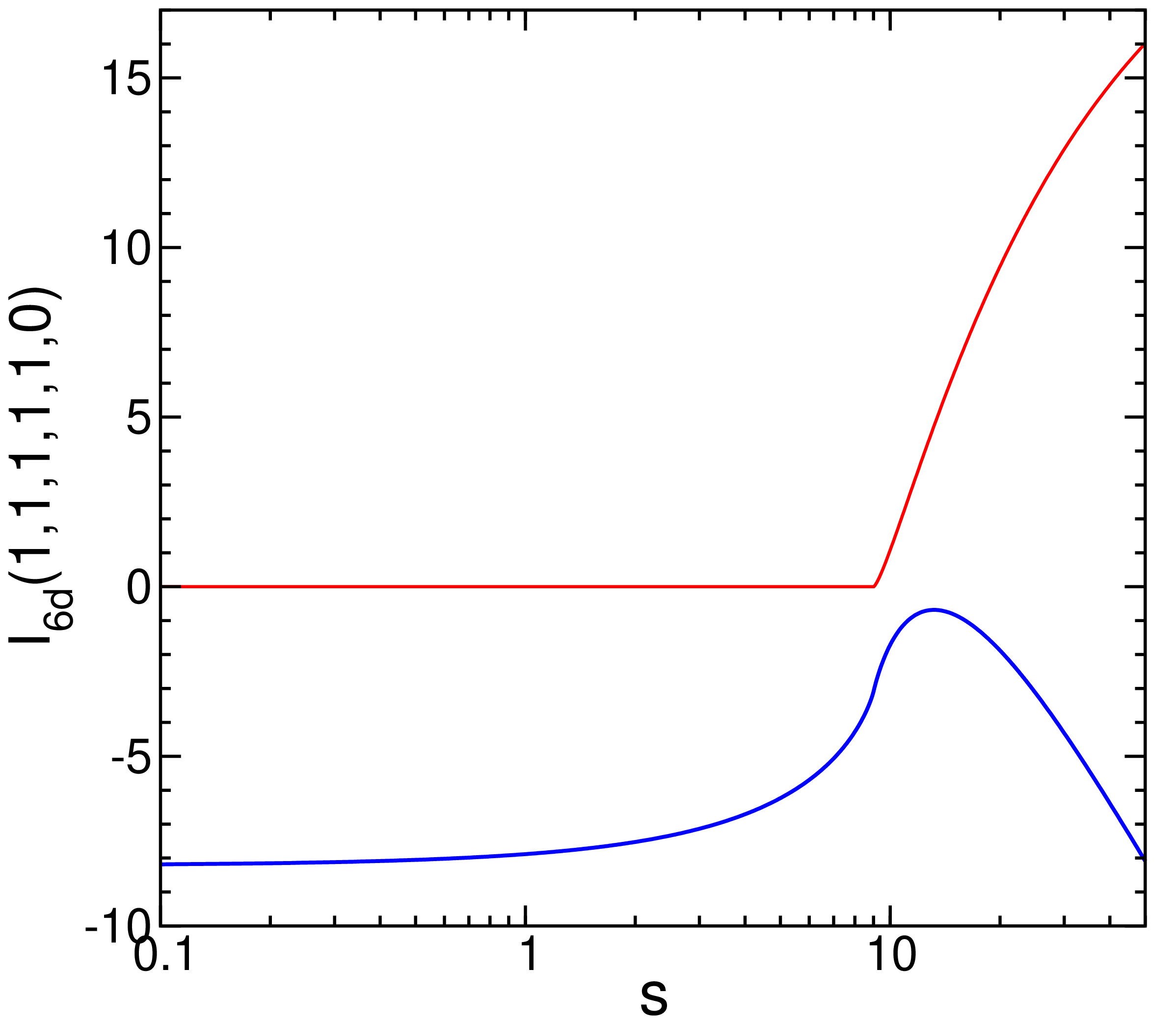}
  \end{minipage}
  \begin{center}
    \begin{minipage}[]{0.92\linewidth}
\caption{\label{fig:integraldataodd}Sample results, for some dimensionless (6-propagator) odd-threshold integrals, with all internal propagator squared masses and the renormalization scale $Q$ set equal to unity ($t= Q^2 = 1$), as a function of the external momentum invariant $s$. The results were obtained by numerical solution of the coupled first-order differential equations in $s$ as provided in the ancillary file {\tt Ioddsdds}, starting from the series solution provided in the ancillary file {\tt Ioddseries} evaluated at $s_0=0.5$ as the initial condition. In each case, the blue (heavier) line is the real part, and the red (lighter) line is the imaginary part. The lowest threshold is at $s=1$ for $I_{6a}(1,0,0,0,1,1)$ (with a logarithmic singularity)  and $I_{6e}(0,0,1,1,1,1)$ and $I_{6d}(1,0,0,0,0,0)$, with the first two also having visibly conspicuous thresholds at $s=9$. For $I_{6d}(1,1,1,1,1,0)$, the only threshold is at $s=9$.}
    \end{minipage}
  \end{center}
\end{figure}

For larger values of $s$, as explained in section \ref{subsec:numericaleval}, it is straightforward to numerically integrate the first-order coupled linear differential equations for the master integrals as a function of the dependent variable $s$, starting from, for example, $s_0=0.5 t$ where the numerical values can be obtained using the series solution. A few examples of results for some dimensionless (6-propagator) integrals as a function of $s$ are shown in Figure \ref{fig:integraldataodd}, for fixed $t = Q^2 = 1$.
Here, $I_{6a}(1,0,0,0,1,1)$ has a threshold at $s=1$, with a logarithmic singularity. The integrals $I_{6e}(0,0,1,1,1,1)$ and $I_{6d}(1,0,0,0,0,0)$ remain finite at the thresholds at $s=1$. Both $I_{6a}(1,0,0,0,1,1)$ and $I_{6e}(0,0,1,1,1,1)$ also have visibly conspicuous three-particle-cut thresholds at $s=9$. For $I_{6d}(1,1,1,1,1,0)$, the only threshold is  at $s=9$.
     
\section{Integrals with even thresholds \label{sec:even}}
\setcounter{equation}{0}
\setcounter{figure}{0}
\setcounter{table}{0} 
\setcounter{footnote}{1}

\baselineskip=15.6pt

\begin{figure}[!t]
\begin{center}
\includegraphics[width=14cm,angle=0]{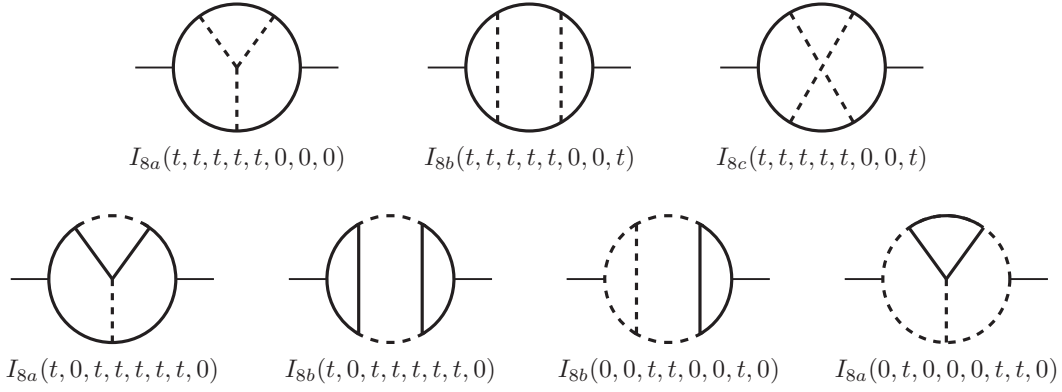}
\end{center}
\vspace{-0.3cm}
\begin{minipage}[]{0.95\linewidth}
\caption{\label{fig:eventhresholdtopologies}
The even-threshold single-mass three-loop self-energy topologies considered (along with their descendants) in section \ref{sec:even}. The heavy solid internal lines represent propagators with squared mass $t$, and the dashed lines represent massless propagators.}
\end{minipage}
\end{figure}

Consider the three-loop self energy integrals corresponding to the topologies shown in Figure~\ref{fig:eventhresholdtopologies}: 
\beq
&&I_{8a}(t,t,t,t,t,0,0,0),\>\>\> I_{8b}(t,t,t,t,t,0,0,t),\>\>\> I_{8c}(t,t,t,t,t,0,0,t),
\nonumber \\ &&
I_{8a}(t,0,t,t,t,t,t,0),\>\>\>I_{8b}(t,0,t,t,t,t,t,0),\>\>\> I_{8b}(0,0,t,t,0,0,t,0),\>\>\> I_{8a}(0,t,0,0,0,t,t,0),
\phantom{xxx}
\eeq
and their descendants obtained by removing internal lines in all possible ways. Note that these 8-propagator topologies are linked by various common descendants. They have possible thresholds at $s=0$, $s=4t$, and/or $s=16 t$, but never at $s=t$ or $s=9t$, and so are referred to here as ``even threshold" integrals. These, along with the all-massless integrals of section \ref{sec:zeromass}, arise in QCD corrections to the self-energies of the $Z$ and Higgs bosons. The QCD corrections to the
$W$ self-energy in which the $W$ couples to massless quarks also include
descendants of $I_{8a}(0,t,0,0,0,t,t,0)$.

Applying the IBP identities, I find that the resulting renormalized $\epsilon$-finite masters at one-loop, two-loop, and three-loop orders can be chosen as (omitting the analytically known $B(0,0)$ and $M(0,0,0,0,0)$, found in section \ref{sec:zeromass}):
\beq
{\cal I}_1 &=&
\{
A(t),\> 
B(t, t)
\}
,
\label{eq:eventhreshbasisone}
\\ 
{\cal I}_2 &=&
\{
V(t, t, 0, t),\> 
M(0, t, 0, t, t),\>
M(t, t, t, t, 0) 
\}
,
\\
{\cal I}_3 &=& \{
H(0, 0, t, 0, t, t),\> 
H(0, t, t, t, 0, t),\>     
I_4(t, t, t, t),\> 
I_{5a}(t, 0, t, 0, t),\>
I_{5b}(0, t, t, t, t),\> 
\nonumber \\ &&
I_{5c}(t, t, t, t, t),\> 
I_{6c}(t, t, t, 0, t, t),\>
I_{6c2}(t, t, t, 0, 0, 0),\> 
I_{6d}(0, 0, 0, t, t, t),\> 
I_{6d}(0, t, t, t, t, 0),\>
\nonumber \\ &&
I_{6d}(t, 0, t, 0, t, 0),\> 
I_{6d}(t, 0, t, t, 0, t),\> 
I_{6e}(0, 0, 0, 0, t, t),\>
I_{6e}(0, t, t, t, 0, t),\> 
I_{6e}(t, t, t, 0, t, t),\>
\nonumber \\ && 
I_{6e1}(t, t, t, 0, 0, 0),\>
I_{6f}(0, 0, 0, 0, t, t),\>
I_{6f5}(0, 0, 0, 0, t, t),\> 
I_{7a}(0, 0, t, t, t, t, t),\>
\nonumber \\ && 
I_{7a}(t, t, 0, 0, 0, 0, t),\>
I_{7a}(0, t, 0, t, 0, t, 0),\>
I_{7a3}(t, 0, t, 0, t, 0, 0),\>
I_{7a}(t, t, t, t, t, t, 0),\>
\nonumber \\ &&  
I_{7a3}(t, t, t, t, t, t, 0),\> 
I_{7b}(0, t, t, t, t, 0, 0),\>
I_{7b}(t, 0, t, 0, 0, 0, t),\>
I_{7b}(t, 0, t, t, t, t, 0),\>
\nonumber \\ &&  
I_{7b4}(t, 0, t, t, t, t, 0),\> 
I_{7b4}(t, t, 0, t, t, 0, t),\>
I_{7c}(t, t, t, t, 0, 0, 0),\> 
I_{7d}(t, t, 0, t, 0, t, 0),\>
\nonumber \\ &&  
I_{7d}(t, t, 0, t, t, 0, t),\> 
I_{7e}(0, 0, 0, 0, 0, t, t),\>
I_{7e}(0, 0, t, t, t, 0, 0),\>
I_{8a}(t, 0, t, t, t, t, t, 0),\> 
\nonumber \\ &&  
I_{8a}(t, t, t, t, t, 0, 0, 0),\>
I_{8b}(t, 0, t, t, t, t, t, 0),\>
I_{8b}(0, 0, t, t, 0, 0, t, 0),\>
I_{8b}(t, t, t, t, t, 0, 0, t),\>
\nonumber \\ &&  
I_{8c}(t, 0, t, t, t, t, t, 0),\> 
I_{8c}^{pk}(t, t, t, t, t, 0, 0, t)     
\}
.
\label{eq:eventhreshbasisthree}
\eeq
For the other candidate master integrals, the solutions (obtained from the IBP relations) in terms of the masters are given in an 
ancillary file {\tt Ieven}. The complete list of such solved $\epsilon$-finite candidate master integrals is (omitting the analytically known massless integrals $S(0,0,0)$ and $U(0,0,0,0)$, found in section \ref{sec:zeromass}):
\beq
&&\{
I(0, t, t),\> 
S(0, t, t),\> 
T(t, 0, t),\> 
U(0, 0, t, t),\> 
U(t, t, 0, t),\>
F(t, 0, 0, t),\> 
F(t, t, t, t),\> 
\nonumber \\ &&  
G(0, 0, 0, t, t),\> 
G(0, t, t, t, t),\> 
G(t, 0, t, 0, t),\> 
I_{4}(0, 0, t, t),\> 
I_{41}(t, 0, 0, t),\> 
I_{41}(t, t, t, t),\> 
\nonumber \\ &&  
I_{411}(t, 0, 0, t),\>
I_{411}(t, t, t, t),\>
I_{412}(t, t, 0, 0),\>
I_{412}(t, t, t, t),\>
I_{5a}(0, 0, 0, t, t),\>
I_{5a}(0, t, t, t, t),\>
\nonumber \\ &&  
I_{5a1}(t, 0, t, 0, t),\>
I_{5a2}(0, t, t, 0, 0),\>
I_{5a2}(0, t, t, t, t),\>
I_{5a2}(t, t, 0, 0, t),\>
I_{5b}(0, 0, 0, t, t),\>
\nonumber \\ &&  
I_{5b}(0, t, t, 0, 0),\>
I_{5b}(t, 0, t, 0, t),\>
I_{5b1}(t, 0, t, 0, t),\>
I_{5b2}(0, t, t, 0, 0),\>
I_{5b2}(0, t, t, t, t),\>
\nonumber \\ &&  
I_{5b2}(t, t, 0, 0, t),\>
I_{5b4}(0, 0, 0, t, t),\>
I_{5b4}(0, t, t, t, t),\>
I_{5b4}(t, 0, t, t, 0),\>
I_{5c}(0, 0, 0, t, t),\>
\nonumber \\ &&  
I_{5c}(t, t, 0, 0, t),\>
I_{5c3}(0, 0, t, 0, t),\>
I_{5c3}(t, t, t, 0, 0),\>
I_{5c3}(t, t, t, t, t),\>
I_{5c2}(t, t, 0, 0, t),\>
\nonumber \\ &&  
I_{5c2}(t, t, t, t, t),\>
I_{5c1}(t, t, 0, 0, t),\>
I_{5c1}(t, t, t, t, t),\>
I_{6a}(0, 0, 0, 0, t, t),\>
I_{6a}(0, 0, t, t, t, t),\>
\nonumber \\ &&  
I_{6a}(t, t, 0, t, 0, t),\>
I_{6a1}(t, t, 0, t, 0, t),\>
I_{6a2}(t, t, 0, t, 0, t),\>
I_{6a3}(0, 0, t, t, 0, 0),\>
I_{6a3}(0, 0, t, t, t, t),\>
\nonumber \\ &&  
I_{6a3}(t, t, t, 0, 0, t),\>
I_{6b}(0, 0, 0, 0, t, t),\>
I_{6b}(0, 0, t, t, t, t),\>
I_{6b}(t, t, 0, t, 0, t),\>
I_{6b1}(t, t, 0, t, 0, t),\>
\nonumber \\ &&  
I_{6b3}(0, 0, t, t, 0, 0),\>
I_{6b3}(0, 0, t, t, t, t),\>
I_{6b3}(t, t, t, 0, 0, t),\>
I_{6c}(0, 0, 0, 0, t, t),\>
I_{6c}(0, 0, t, t, 0, t),\>
\nonumber \\ &&  
I_{6c}(t, t, 0, t, 0, t),\>
I_{6c}(t, t, t, 0, 0, 0),\>
I_{6c1}(t, t, 0, t, 0, t),\>
I_{6c1}(t, t, t, 0, 0, 0),\>
I_{6c1}(t, t, t, 0, t, t),\>
\nonumber \\ &&  
I_{6c2}(t, t, 0, t, 0, t),\>
I_{6c2}(t, t, t, 0, t, t),\>
I_{6c3}(0, 0, t, t, 0, t),\>
I_{6c3}(t, t, t, 0, 0, 0),\>
I_{6c3}(t, t, t, 0, t, t),\>
\nonumber \\ &&  
I_{6c4}(0, 0, t, t, 0, t),\>
I_{6c4}(t, t, 0, t, 0, t),\>
I_{6c5}(0, 0, 0, 0, t, t),\>
I_{6c5}(0, 0, t, t, t, 0),\>
I_{6c5}(t, t, 0, t, t, 0),\>
\nonumber \\ &&  
I_{6c5}(t, t, t, 0, t, t),\>
I_{6d1}(t, 0, t, 0, t, 0),\>
I_{6d1}(t, 0, t, t, 0, t),\>
I_{6d2}(0, t, t, 0, 0, t),\>
I_{6d2}(0, t, t, t, t, 0),\>
\nonumber \\ &&  
I_{6d2}(t, t, 0, 0, t, t),\>
I_{6d2}(t, t, 0, t, 0, 0),\>
I_{6d6}(0, 0, 0, t, t, t),\>
I_{6d6}(t, 0, t, t, 0, t),\>
I_{6e}(t, 0, 0, t, 0, t),\>
\nonumber \\ &&  
I_{6e}(t, t, t, 0, 0, 0),\>
I_{6e1}(t, 0, 0, t, 0, t),\>
I_{6e1}(t, t, t, 0, t, t),\>
I_{6e5}(0, 0, 0, 0, t, t),\>
I_{6e5}(0, t, t, t, t, 0),\>
\nonumber \\ &&  
I_{6e5}(t, 0, 0, t, t, 0),\>
I_{6e5}(t, t, t, 0, t, t),\>
I_{6e2}(0, t, t, t, 0, t),\>
I_{6e2}(t, t, t, 0, 0, 0),\>
I_{6e2}(t, t, t, 0, t, t),\>
\nonumber \\ &&  
I_{6e4}(0, t, t, t, 0, t),\>
I_{6e4}(t, 0, 0, t, 0, t),\>
I_{6e3}(0, t, t, t, 0, t),\>
I_{6e3}(t, t, t, 0, 0, 0),\>
I_{6e3}(t, t, t, 0, t, t),\>
\nonumber \\ &&  
I_{6f}(0, 0, t, t, 0, t),\>
I_{6f}(t, t, t, t, 0, 0),\>
I_{6f}(t, t, t, t, t, t),\>
I_{6f5}(0, 0, t, t, t, 0),\>
I_{6f5}(t, t, t, t, t, t),\>
\nonumber \\ &&  
I_{6f1}(t, t, 0, 0, 0, t),\>
I_{6f1}(t, t, t, t, 0, 0),\>
I_{6f1}(t, t, t, t, t, t),\>
I_{7a1}(t, 0, t, 0, t, 0, 0),\>
\nonumber \\ &&  
I_{7a1}(t, t, 0, 0, 0, 0, t),\>
I_{7a1}(t, t, t, t, t, t, 0),\>
I_{7a3}(0, 0, t, t, t, t, t),\>
I_{7a5}(0, 0, t, t, t, t, t),\>
\nonumber \\ &&  
I_{7a5}(t, 0, t, 0, t, 0, 0),\>
I_{7a5}(t, t, t, t, t, t, 0),\>
I_{7a7}(0, 0, t, t, t, t, t),\>
I_{7a7}(t, t, 0, 0, 0, 0, t),\>
\nonumber \\ &&  
I_{7b1}(t, 0, t, 0, 0, 0, t),\>
I_{7b1}(t, 0, t, t, t, t, 0),\>
I_{7b2}(0, t, t, t, t, 0, 0),\>
I_{7b2}(t, t, 0, 0, 0, t, 0),\>
\nonumber \\ &&  
I_{7b2}(t, t, 0, t, t, 0, t),\>
I_{7b4}(0, t, t, t, t, 0, 0),\>
I_{7b6}(t, 0, t, t, t, t, 0),\>
I_{7b6}(t, t, 0, 0, 0, t, 0),\>
\nonumber \\ &&  
I_{7c}(0, 0, 0, 0, 0, t, t),\>
I_{7c}(t, t, t, t, 0, t, t),\>
I_{7c1}(t, t, t, t, 0, 0, 0),\>
I_{7c1}(t, t, t, t, 0, t, t),\>
\nonumber \\ &&  
I_{7c6}(0, 0, 0, 0, 0, t, t),\>
I_{7c6}(t, t, t, t, 0, t, t),\>
I_{7d}(0, 0, 0, 0, t, t, t),\>
I_{7d1}(t, t, 0, t, 0, t, 0),\>
\nonumber \\ &&  
I_{7d1}(t, t, 0, t, t, 0, t),\>
I_{7d2}(t, t, 0, t, 0, t, 0),\>
I_{7d2}(t, t, 0, t, t, 0, t),\>
I_{7d3}(0, 0, t, t, 0, 0, t),\>
\nonumber \\ &&  
I_{7d3}(t, t, t, 0, 0, t, t),\>
I_{7d3}(t, t, t, 0, t, 0, 0),\>
I_{7d7}(0, 0, 0, 0, t, t, t),\>
I_{7d7}(t, t, 0, t, t, 0, t),\>
\nonumber \\ &&  
I_{7e}(0, 0, t, t, t, t, t),\>
I_{7e}(t, t, t, t, 0, 0, t),\>
I_{7e1}(t, t, t, t, 0, 0, t),\>
I_{7e2}(t, t, t, t, 0, 0, t),\>
\nonumber \\ &&  
I_{7e3}(0, 0, t, t, t, 0, 0),\>
I_{7e3}(0, 0, t, t, t, t, t),\>
I_{7e3}(t, t, t, t, 0, 0, t),\>
I_{7e4}(0, 0, t, t, t, 0, 0),\>
\nonumber \\ &&  
I_{7e4}(0, 0, t, t, t, t, t),\>
I_{7e4}(t, t, t, t, 0, 0, t),\>
I_{7e5}(0, 0, t, t, t, 0, 0),\>
I_{7e5}(0, 0, t, t, t, t, t),\>
\nonumber \\ &&  
I_{7e6}(0, 0, 0, 0, 0, t, t),\>
I_{7e6}(0, 0, t, t, t, t, t),\>
I_{7e6}(t, t, t, t, 0, t, 0),\>
I_{8a}(0, t, 0, 0, 0, t, t, 0)
\}
.
\label{eq:evenbasissolved}
\eeq

Furthermore, the derivatives with respect to $s$ of the master integrals in eqs.~(\ref{eq:eventhreshbasisone})-(\ref{eq:eventhreshbasisthree}) can be re-expressed in terms of the same set of masters. 
The results for $\displaystyle{s \frac{d}{ds}}$ acting on all of the master
integrals listed are provided in another ancillary file {\tt Ievensdds}. 

For numerical evaluation, I have derived the series solutions to the differential equations in $s$, using as boundary conditions the values of the integrals at $s=0$, which can be obtained from the results for the vacuum integrals, including eq.~(\ref{eq:H0ttt0t}) and 
\beq
H(0,0,t,0,t,t) &=& -9 \zeta_4 + 6 \zeta_3 (1 - L_t),
\eeq
using the notation of ref.~\cite{Martin:2016bgz}.
The series results, up to order $r^{36}$, for all of the integrals listed in eqs.~(\ref{eq:eventhreshbasisone})-(\ref{eq:eventhreshbasisthree}) and (\ref{eq:evenbasissolved}) 
are given in the ancillary file 
{\tt Ievenseries},
in terms of $r = s/t$, $L_t = \lnbar(t)$, and
$L_{-s} = \lnbar(s) - i \pi$ 
and the constants $\zeta_3$, $\zeta_4$, $\conHfour$,
and other coefficients that are rational numbers.
These series solutions converge for $|r| < 4$, which is sufficient for evaluating
the three-loop leading QCD corrections to the Higgs and $Z$ boson pole masses in the Standard Model with $r = m_Z^2/m_t^2$.

\begin{figure}[!t]
  \begin{minipage}[]{0.495\linewidth}
    \includegraphics[width=7.cm,angle=0]{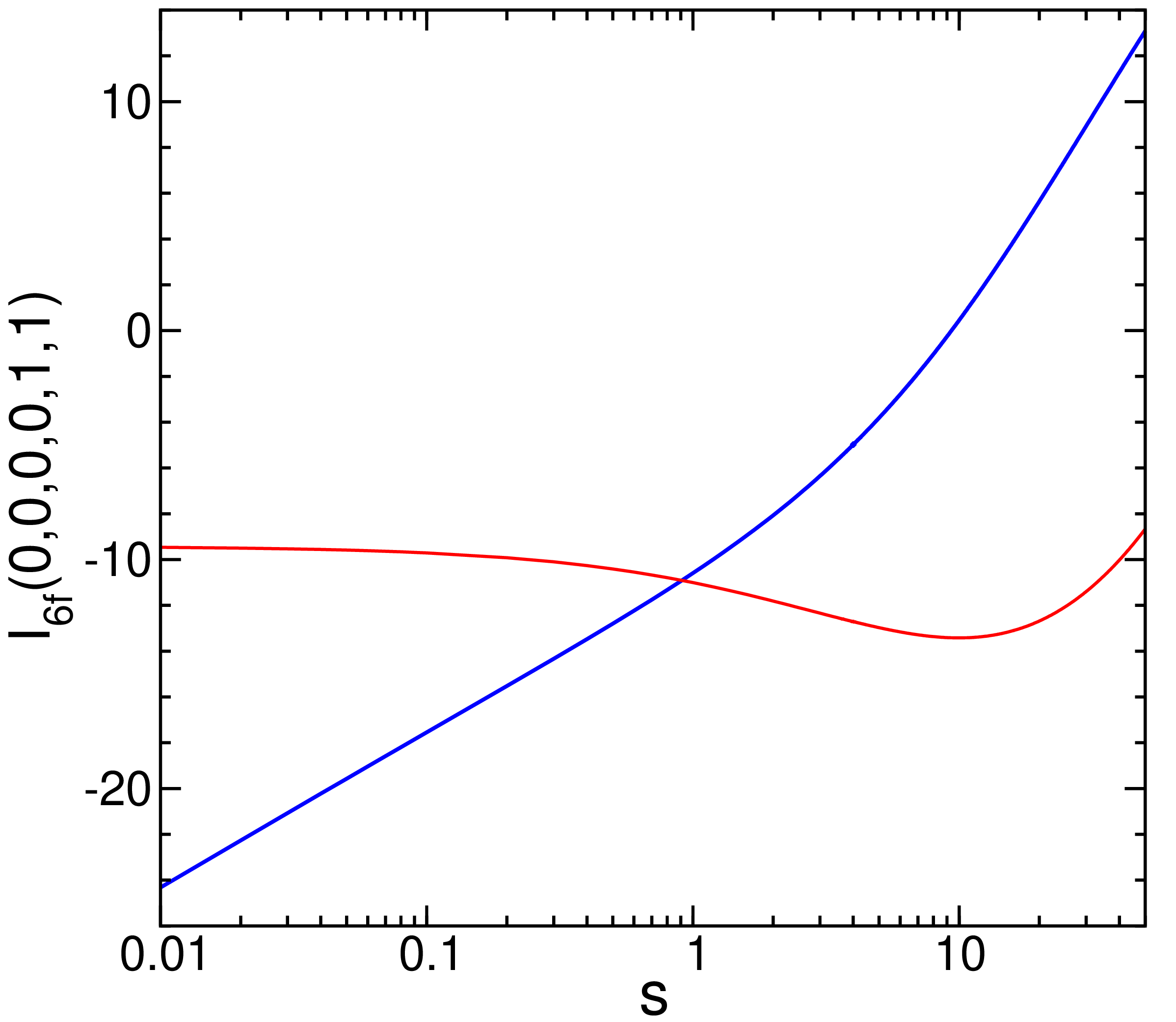}
  \end{minipage}
  \begin{minipage}[]{0.495\linewidth}
    \includegraphics[width=7.cm,angle=0]{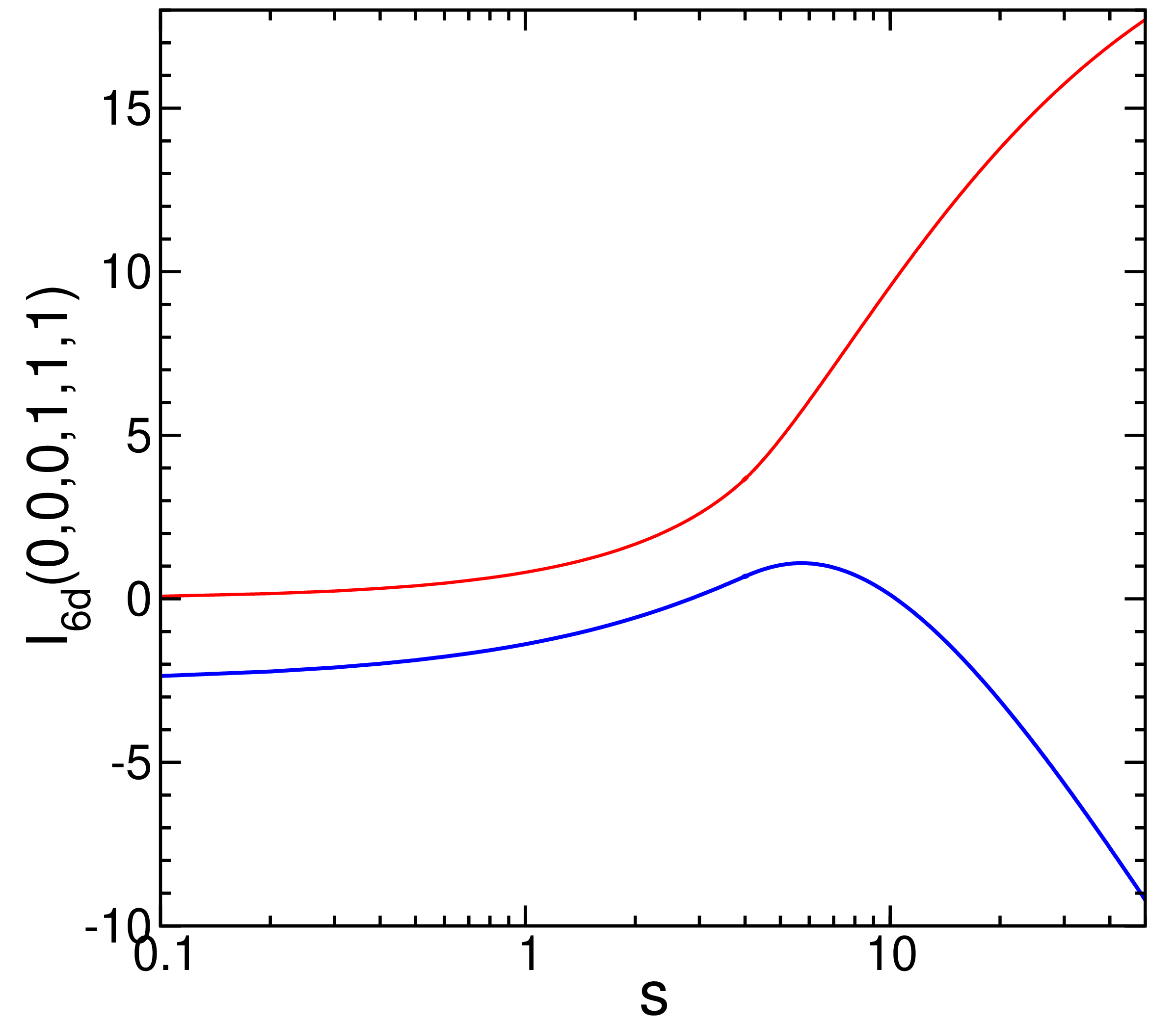}
  \end{minipage}
  \begin{minipage}[]{0.495\linewidth}
    \includegraphics[width=7.cm,angle=0]{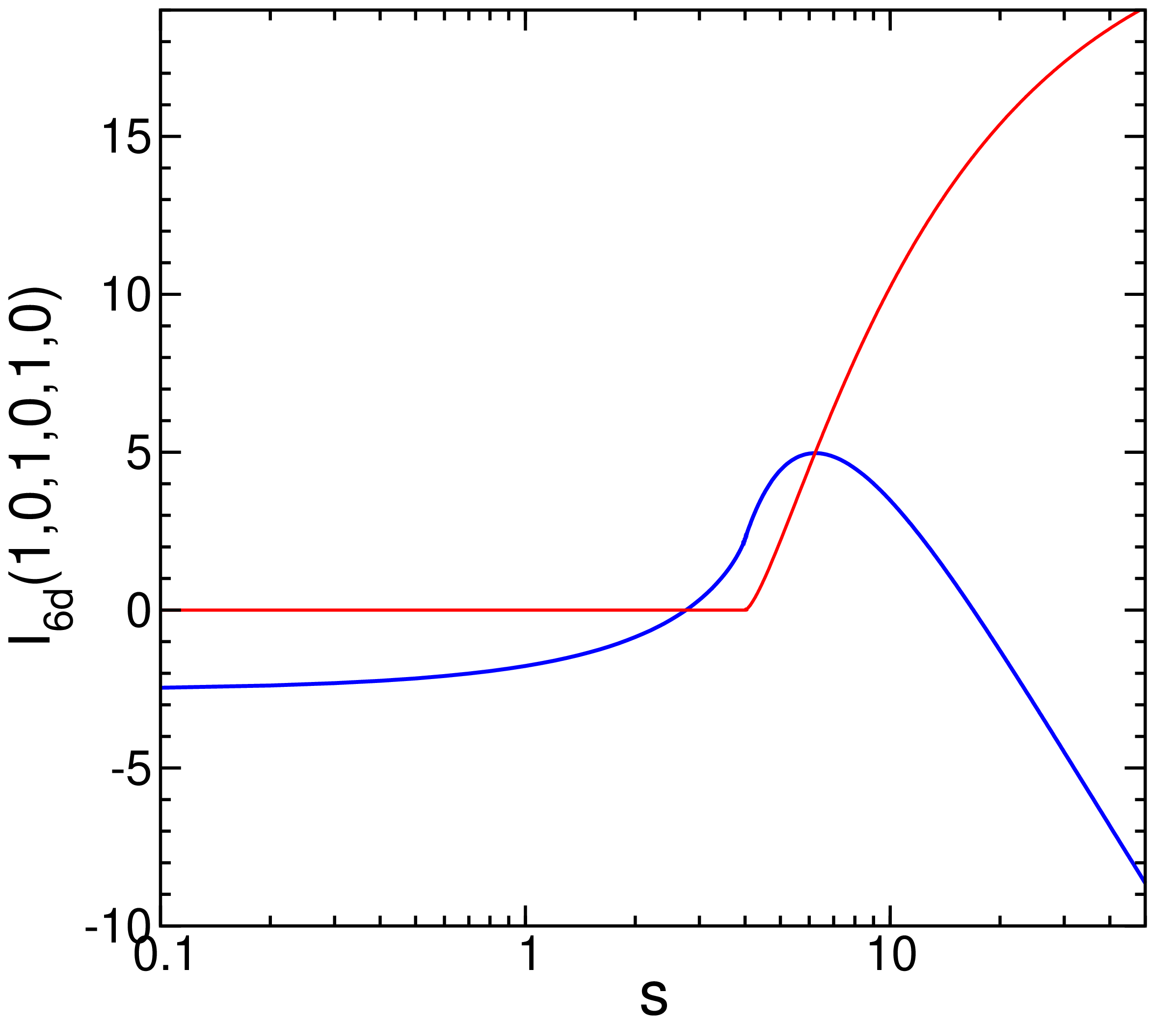}
  \end{minipage}
   \begin{minipage}[]{0.495\linewidth}
    \includegraphics[width=7.cm,angle=0]{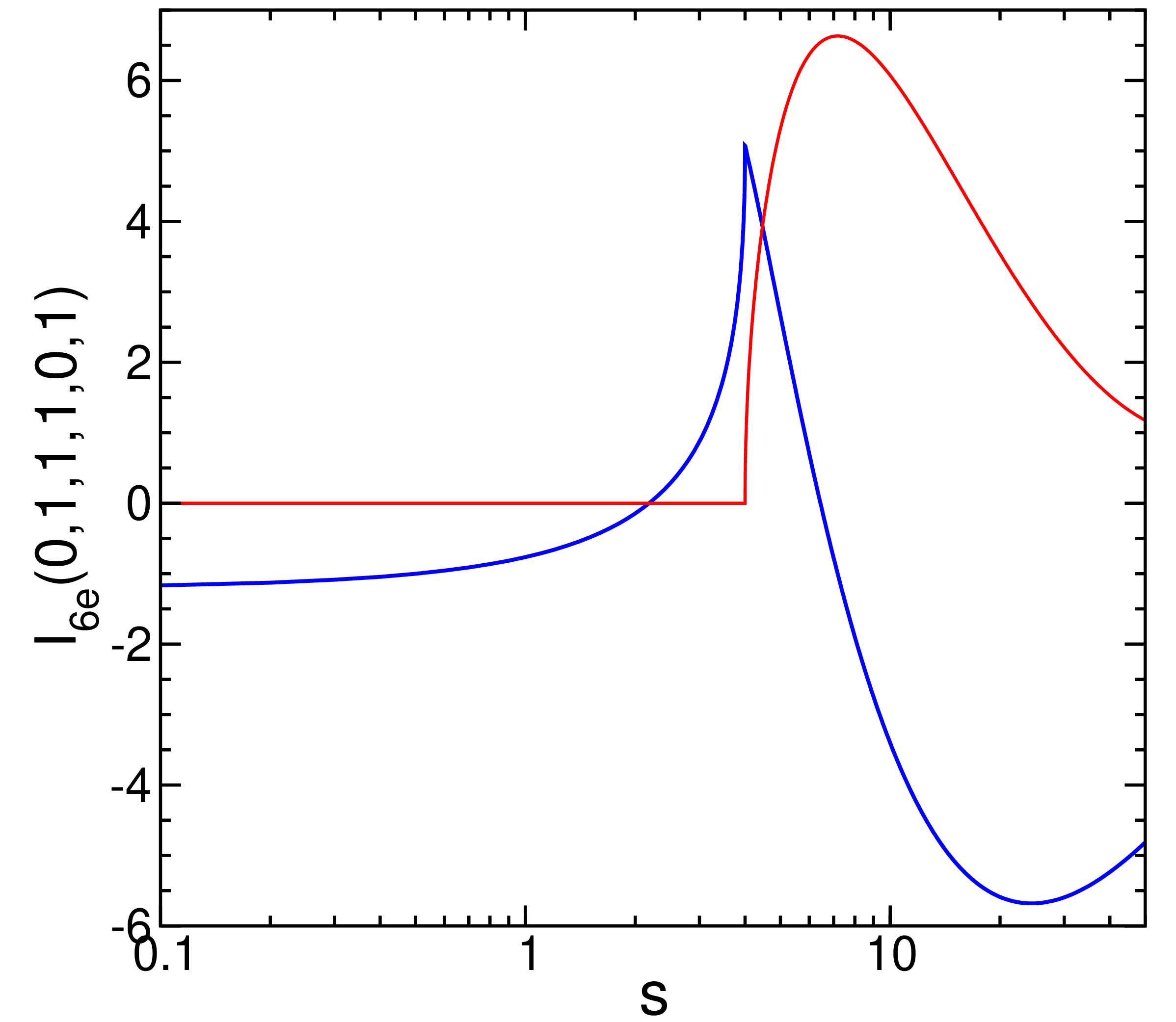}
  \end{minipage}
  \begin{center}
\begin{minipage}[]{0.95\linewidth}
\vspace{-0.3cm}
\caption{\label{fig:integraldataeven}Sample results, for some dimensionless (6-propagator) even-threshold integrals, with all internal propagator squared masses and the renormalization scale $Q$ set equal to unity ($t= Q^2 = 1$), as a function of the external momentum invariant $s$. The results were obtained by  numerical solution of the coupled first-order differential equations in $s$ as provided in the ancillary file {\tt Ievensdds}, starting from the series solution provided in the ancillary file {\tt Ievenseries} evaluated at $s_0=0.5$ as the initial condition. In each case, the blue (heavier) line is the real part, and the red (lighter) line is the imaginary part. The lowest threshold is at $s=0$ (two-particle cut, with logarithmic singularity) for $I_{6f}(0,0,0,0,1,1)$, at $s=0$ (three-particle cut) for $I_{6d}(0,0,0,1,1,1)$, and at $s=4$ for $I_{6d}(1,0,1,0,1,0)$ and $I_{6e}(0,1,1,1,0,1)$.}
    \end{minipage}
  \end{center}
\end{figure}

As explained in section \ref{subsec:numericaleval}, for larger values of $s$, one can numerically integrate the first-order coupled linear differential equations for the master integrals as a function of the dependent variable $s$, starting from e.g.~$s_0=0.5 t$ where the numerical values can be obtained using the series solutions. 
As one numerical consistency check, I have verified that the results for $s \gg t$ reproduce those given in eqs.~(\ref{eq:Ballmassless})-(\ref{eq:I8cpkallmassless}). 
Some examples of results for dimensionless (6-propagator) integrals as a function of $s$ are shown in Figure \ref{fig:integraldataeven}, for fixed $t = Q^2 = 1$.
The function $I_{6f}(0,0,0,0,1,1)$ has a threshold due to a two-particle cut at $s=0$, with a logarithmic singularity there. The function $I_{6d}(0,0,0,1,1,1)$ has a three-particle cut threshold at $s=0$, with no singularity. The functions  $I_{6d}(1,0,1,0,1,0)$ and $I_{6e}(0,1,1,1,0,1)$ have their lowest thresholds at $s=4$, where the latter has a sharp cusp but remains finite.

\section{Outlook\label{sec:outlook}}
\setcounter{equation}{0}
\setcounter{figure}{0}
\setcounter{table}{0} 
\setcounter{footnote}{1}

In this paper, I have formalized the concept of renormalized $\epsilon$-finite master integrals, in which UV sub-divergences are subtracted. These have the advantage that the expansions of the master integrals to positive powers of $\epsilon$ never appear. (One hand-wavy way of understanding why this is not totally unexpected is that 
the calculations of renormalized observables could in principle employ some other regulator, not based on dimensional continuation at all, in which case there would be no reason for the expansions of the integrals for finite $\epsilon$.) The necessary subtractions were given explicitly for three-loop self-energy integrals in section \ref{subsec:threeSE}. 

I also carried out the solution of the IBP relations for the cases with one internal mass scale (and some vanishing propagator masses), and provided the results needed for fast and accurate numerical evaluation
of the renormalized $\epsilon$-finite masters. The results obtained here are applied to the calculation of the 3-loop QCD corrections to the physical masses of the Standard Model $W$, $Z$, and Higgs bosons in the pure \MSbar tadpole-free scheme in ref.~\cite{SPMtoappear}. The same methods can be applied to numerically calculate three-loop self-energy integrals for arbitrary masses, although the coefficients will be significantly more complicated in the general case, and the number of distinct master integrals will of course be much larger.

In this paper, I have not attempted to specifically address situations with more than two external legs. In that case, non-UV singular contributions involving virtual massless particles require cancellation with the contributions from real emission diagrams at lower loop order, but the same principle of incorporating the UV counterterms within the $\epsilon$-finite master integrals should be beneficial.

{\it Acknowledgments:} This work was supported in part by the National Science 
Foundation grant number 2013340. 


\end{document}